\documentclass[acmtog]{acmart}
\AtBeginDocument{%
  \providecommand\BibTeX{{%
    \normalfont B\kern-0.5em{\scshape i\kern-0.25em b}\kern-0.8em\TeX}}}

\setcopyright{rightsretained}
\acmJournal{TOG}
\acmYear{2023} \acmVolume{1} \acmNumber{1} \acmArticle{1} \acmMonth{1} \acmPrice{}\acmDOI{10.1145/3592407}

\acmPrice{15.00}
\acmISBN{978-1-4503-XXXX-X/18/06}

\usepackage{xcolor}

\definecolor{dkgreen}{rgb}{0.05, 0.5, 0.06}
\definecolor{mauve}{rgb}{0.88, 0.69, 1.0}

\usepackage{listings}
\usepackage{adjustbox}
\usepackage{xspace}
\usepackage{pgfplots}
\usepackage{tabularx}
\usepackage{xcolor}
\usepackage{pgf}
\usepackage{multirow}
\usepackage{colortbl}
\usepackage{booktabs}
\usepackage{enumitem}

\definecolor{Gray}{gray}{0.925}
\newcolumntype{g}{>{\columncolor{Gray}}c}

\newcommand{\papertitle}{Random-Access Neural Compression of Material Textures}

\newcommand{\FLIP}{\protect\reflectbox{F}LIP\xspace}
\newcommand{\tabvspace}{-3.0mm}

 \usepackage{hyphenat}

\begin{document}

\title{\papertitle} %

\author{Karthik Vaidyanathan}
\authornote{Authors contributed equally to this work.}
\email{kvaidyanatha@nvidia.com}
\affiliation{
   \institution{NVIDIA}
   \country{USA}
}
\author{Marco Salvi}
\authornotemark[1]
\email{msalvi@nvidia.com}
\affiliation{
   \institution{NVIDIA}
   \country{USA}
}
\author{Bartlomiej Wronski}
\authornotemark[1]
\email{bwronski@nvidia.com}
\affiliation{
   \institution{NVIDIA}
   \country{USA}
}
\author{Tomas Akenine-M\"{o}ller}
\email{takenine@nvidia.com}
\affiliation{
   \institution{NVIDIA}
   \country{Sweden}
}
\author{Pontus Ebelin}
\email{pandersson@nvidia.com}
\affiliation{
   \institution{NVIDIA}
   \country{Sweden}
}
\author{Aaron Lefohn}
\email{alefohn@nvidia.com}
\affiliation{
   \institution{NVIDIA}
   \country{USA}
}

\begin{abstract}

The continuous advancement of photorealism in rendering is accompanied by a growth in texture data and, consequently, increasing storage and memory demands. To address this issue, we propose a novel neural compression technique specifically designed for material textures.
We unlock two more levels of detail, i.e., 16$\times$ more texels, using low bitrate compression, with image quality that is better than advanced image compression techniques, such as AVIF and JPEG XL. 

At the same time, our method allows on-demand, real-time decompression with random access similar to block texture compression on GPUs, enabling compression on disk and memory. 
The key idea behind our approach is compressing multiple material textures and their mipmap chains together, and using a small neural network, that is optimized for each material, to decompress them. Finally, we use a custom training implementation to achieve practical compression speeds, whose performance surpasses that of general frameworks, like PyTorch, by an order of magnitude.

\end{abstract}



\begin{teaserfigure}
	\centering
	\newcommand{\imgH}{0.214\textwidth}
	\newcommand{\imgHH}{0.1569\textwidth}
	\setlength{\tabcolsep}{1.0pt}%
	\newcommand{\linelen}{23.5mm}
	{\scriptsize
		\begin{tabular}{ccccc}
			& \textbf{BC high}. PSNR (\textuparrow): 19.4~dB, \FLIP (\textdownarrow): 0.224   
			& \textbf{NTC}. PSNR (\textuparrow): \textbf{22.0}~dB, \FLIP (\textdownarrow): \textbf{0.177}
			& \textbf{reference}: not compressed
                \\
                & $1024\times 1024$ at 5.3~MB.
                & $4096\times 4096$ at \textbf{3.8}~MB.
                & $4096\times 4096$ at 256~MB.
                \\
			\hspace{1.0mm}
			\includegraphics[height=\imgH]{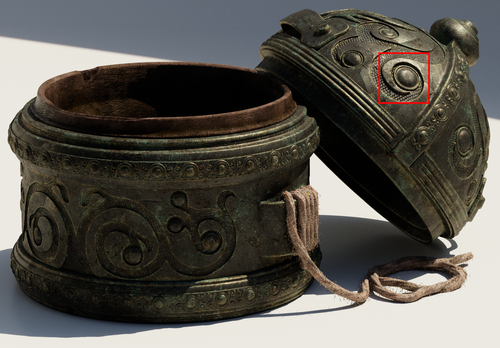} &
			\includegraphics[height=\imgH]{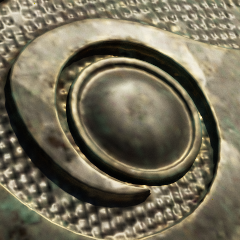} &
			\includegraphics[height=\imgH]{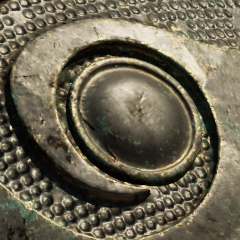} &
			\includegraphics[height=\imgH]{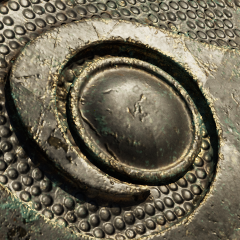}
			\vspace*{-9.7mm}
			\\
			&
			\hspace*{28.8mm}
			\includegraphics[height=0.05\textwidth]{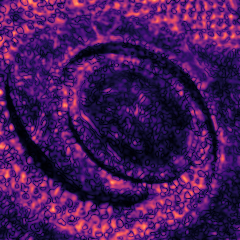}
                &
			\hspace*{28.7mm}
			\includegraphics[height=0.05\textwidth]{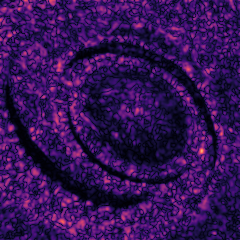} &
			&
			\hspace*{-1.2mm}
			\raisebox{0mm}{\rotatebox{90}{\includegraphics[width=0.05\textwidth, height=1mm]{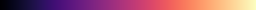}}}
		\end{tabular}
		\hspace*{-0.1mm}
		\begin{tabular}{cccccc}
			\includegraphics[height=\imgHH]{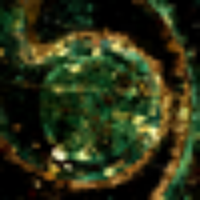} & 
			\includegraphics[height=\imgHH]{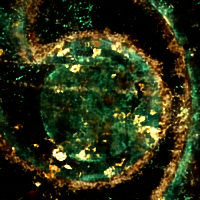} & 
			\includegraphics[height=\imgHH]{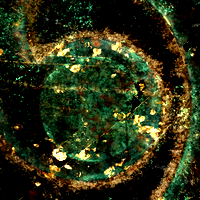} & 
			\includegraphics[height=\imgHH]{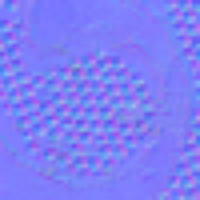} &
			\includegraphics[height=\imgHH]{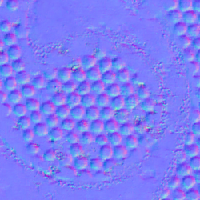} &
			\includegraphics[height=\imgHH]{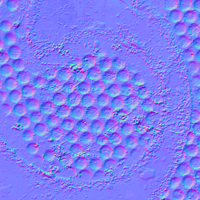}
			\\ 
			BC high & NTC & reference & BC high & NTC & reference
		\end{tabular}
	}
        \vspace{-1em}
        \caption{A rendered image of an inkwell. The cutouts demonstrate quality using, from
        left to right, GPU-based texture formats (BC high) at $1024\times 1024$ resolution,
        our neural texture compression (NTC), and high-quality reference textures. 
        Note that NTC provides a $4\times$ higher resolution ($16\times$ texels) than BC high,
        despite using 30\% less memory.
        The PSNR and \FLIP quality metrics, computed for the cutouts, are shown above the respective images.
        The \FLIP error images are shown in the lower right corners, where brightness is proportional to error. 
        Bottom row: two of the textures that were used for the renderings.
        }
	\Description{TODO}
	\label{fig_teaser}
\end{teaserfigure}

\maketitle
\newpage
\vspace*{-2em}

\section{Introduction}

In recent years, the visual quality of real-time rendering has been approaching the levels of VFX and film productions, giving rise to powerful new workflows, like virtual production~\cite{perkins2022}, which are transforming filmmaking.
These improvements in quality have been achieved through the adoption of methods used in cinematic rendering 
such as physically-based shading for photorealistic modeling of materials~\cite{burley2012physically}, 
ray tracing~\cite{ouyang2021restir,advances2022} and denoising~\cite{hasselgren2020,thomas2022} for accurate global illumination, and technologies like Nanite ~\cite{advances2021} that render compressed micropolygons, thus enabling a significant increase in geometric detail.

Although there is a greater convergence of rendering techniques between cinematic and real-time applications, content creation workflows remain largely different. 
In order to limit storage size, games often use specialized practices for texturing models, which can require significant effort, for example, reusing content through instancing, layering tiled materials, or using procedural effects. In spite of these efforts, games typically present blurry, magnified textures close to the camera.  
Furthermore, some of these techniques are not applicable to uniquely parametrized content, such as photogrammetry, usage of which is a growing trend in games today. 
One of the main obstacles for achieving the next level of realism 
in real-time rendering 
is limited disk storage, download bandwidth, and memory size constraints.

While texture storage requirements in real-time applications have increased significantly, texture compression on GPUs has seen relatively little change. 
GPUs still rely on block-based texture compression methods~\cite{S3TC2022,BCCompression2020,Nystad2012}, 
first introduced in the late 1990s.
These methods have efficient hardware implementations and desirable properties like random access and data locality. They can achieve high, near lossless quality, but are designed for moderate compression ratios, typically between 4$\times$ and 8$\times$. They are also limited to a maximum of 4 channels, while the number of material properties in modern real-time renderers commonly exceed this limit, thus requiring multiple textures.  
The main improvement to texture compression in recent years has been meta-compression for reduced disk storage and faster delivery~\cite{Hurlburt2022}, but this requires transcoding to GPU texture compression formats.

On the other hand, the field of natural image compression is making significant strides in the lower bitrate regime.
In recent years, new web image formats have been proposed~\cite{chen2018overview,alakuijala2019jpeg} that significantly improve upon previous standards, like JPEG~\cite{wallace1991jpeg}.
Meanwhile, the scientific community has been developing \textit{neural image compression} methods~\cite{balle2018,mentzer2020_nips,cheng_2020_cvpr}, incorporating non-linear transformations in the form of neural networks, to aid compression and decompression.
These methods significantly improve upon the perceptual quality of compressed images at extremely low bitrates, but typically offer modest distortion metrics improvements. 
They also require large-scale image data sets and expensive training, and are not suitable for real-time rendering due to their lack of important features, such as random access and non-color material properties compression.

In this work, we tackle the problem of reducing texture storage by integrating techniques from GPU texture compression as well as neural image compression and introducing a neural compression technique specifically designed for material textures.

Using this approach we enable low-bitrate compression, unlocking two additional levels of detail (or 16$\times$ more texels) with similar storage requirements as commonly used texture compression techniques. 
In practical terms, this allows a viewer to get very close to an object before losing significant texture detail. Our main contributions are:

\begin{itemize}[leftmargin=1em]
\item A novel approach to texture compression that exploits redundancies spatially, across mipmap levels, and across different material channels. 
By optimizing for reduced distortion at a low bitrate, we can compress two more levels of details in the same storage as
block-compressed textures. 
The resulting texture quality at such aggressively low bitrates is better than or comparable to recent image compression standards like AVIF and JPEG XL, which are not designed for real-time decompression with random access.
\item A novel low-cost decoder architecture that is optimized specifically for each material. This architecture enables
real-time performance for random access and can be integrated into
material shader functions, such as filtering,
to facilitate on-demand decompression.
\item A highly optimized implementation of our compressor,  with fused backpropogation, enabling practical per-material optimization with resolutions up to $8192\times 8192$ (8k). Our compressor can process a 9-channel, 4k material texture set in 1-15 minutes on an NVIDIA RTX 4090 GPU, depending on the desired quality level. 
\end{itemize}

\section{Previous Work}

In this section, we first review traditional texture compression (TC), techniques used
in contemporary GPUs. Subsequently, we present a brief overview of natural image compression that uses entropy coding 
and its recent development based on deep learning.
Lastly, we examine recent advances in neural rendering 
that are closely related to our work.

\subsection{Traditional Texture Compression}
Delp and Mitchell~\cite{Delp1979} introduced \textit{block truncation coding} (BTC), which compresses
gray scale images by storing two 8-bit gray scale values per $4\times 4$ pixels and having
a single bit per pixel to select one of these two gray scale values.
Each pixel is stored using 2 bits per pixel (BPP).
This was modified by 
Campell et al.~\cite{Campbell1986} who used the 8-bit values as indices
into a lookup table of colors, enabling color image compression at 2~BPP. 
Their method is called color cell compression (CCC).
Knittel et al.~\cite{Knittel1996} described hardware for decompressing CCC textures,
which was selected due its random-access nature
and simplicity, which made it affordable and fast in hardware.

The S3 texture compression (S3TC) schemes~\cite{S3TC2022} are clever extensions of the BTC and CCC
and form the basis for most of the TC methods found in DirectX~\cite{BCCompression2020}.
The first method of S3TC, which was later called DXT1 and then renamed to BC1 in DirectX,
stores two colors per $4\times 4$ pixels. These are quantized to $5+6+5$ (RGB) bits.
Two additional colors are created using linear interpolation between the stored colors.
Hence, there is a palette of four colors available per $4\times 4$ pixels and each pixel
then points to one of these using a 2-bit index.

Today, there are seven variants of S3TC in DirectX. These are called BC1-BC7 and handle
alpha, normal maps, high-dynamic range (HDR), and are using either 4 or 8~BPP.
All of these have the random access property, since each block of $4\times 4$ pixels
always are compressed to the same number of bits.

Munkberg et al.~\cite{Munkberg2006a} and Roimela et al.~\cite{Roimela2006}
presented the first TC schemes for HDR textures,
and both were inspired by the previous block-based schemes, but adapted those to HDR.
BC6H is a variant for HDR texture compression in DirectX. 
We omit many other references on this topic, since HDR TC is not our focus.

Fenney~\cite{Fenney2003} presented a different block-based compression scheme called
PowerVR texture compression (PVRTC), which is used on all iOS devices. PVRTC
decompresses two low-resolution images, which are bilinearly magnified, and then uses a per-pixel index to
select a color in between the interpolated colors.

Ericsson texture compression (ETC1), which is part of OpenGL ES, also compresses 
$4\times 4$ pixels at a time but stores only a single base
color, which is then modulated using a trained table of offsets, which is selected per block~\cite{Strom2005}.
ETC2 is backwards compatible with ETC1 by using invalid bit combinations, and improves
image quality~\cite{Strom2007}.
ETC1/ETC2 are available in over 12 billion mobile phones. %
ASTC~\cite{Nystad2012} is currently the most flexible texture compression scheme,
since it supports low-dynamic range, HDR, and 3D textures, with bitrates from $0.89$ to 8~BPP.
This is achieved using larger block sizes, specific color spaces, and efficient  bit allocation.
ASTC is supported on (at least) ARM's GPUs, the most recent Apple GPUs, as well as several desktop GPUs.
However, there is no support in DirectX.

\subsection{Traditional and Neural Image Compression}
Image compression formats that target storage on disk or network transfer have less restrictive constraints than GPU texture compression.
Without the need for random access and strict bounds on hardware complexity, they can utilize global transforms, as well as entropy coding methods to target significantly lower bitrates.

Despite its widespread usage, JPEG~\cite{wallace1991jpeg} has been found to produce noticeable artifacts, such as detail loss, discoloration, and banding, particularly at lower bitrates. This has led to the development of alternative image compression formats, such as AVIF~\cite{chen2018overview} and JPEG XL~\cite{alakuijala2019jpeg}, which incorporate algorithmic advancements and prioritize alignment with human visual perception~\cite{alakuijala2017guetzli}.

The tradeoff between objective distortion and perceptual measurements has been well-studied~\cite{blau2018perception}, particularly in the context of machine learning.
Various methods have been proposed to improve optimization for perceptual qualities, such as using features extracted from a convolutional network~\cite{Zhang2018}, the network structure itself~\cite{ulyanov2018deep}, or another network~\cite{ledig2017photo}.
Non-neural methods have also been developed to localize perceptually-relevant errors for image comparison~\cite{andersson2020flip}.

In recent years, neural image compression methods~\cite{balle2017,theis2017,rippel2017} have emerged as an alternative to traditional formats such as JPEG 2000~\cite{jpeg2000}, offering improved perceptual quality.  These techniques often use encoder-decoder architectures to create an information bottleneck, which is then quantized and entropy-coded based on an entropy model.
Rapid evolution has occurred in this area, with advances such as the provision of sideband information to improve the accuracy of the entropy model 
~\cite{balle2018}, the incorporation of generative models to achieve higher perceptual quality~\cite{mentzer2020_nips}, and more recently the use of attention based networks~\cite{liu2019attention,cheng_2020_cvpr}
which were the first to improve upon both PSNR and perceptual quality over the new VVC-intra~\cite{vvc} standard, which uses traditional compression methods.

\begin{figure}[t]
    \newcommand{\imgW}{0.24\columnwidth}
    \setlength{\tabcolsep}{1.0pt}%
    {\scriptsize
        \begin{tabular}{cccc}
            \includegraphics[width=\imgW]{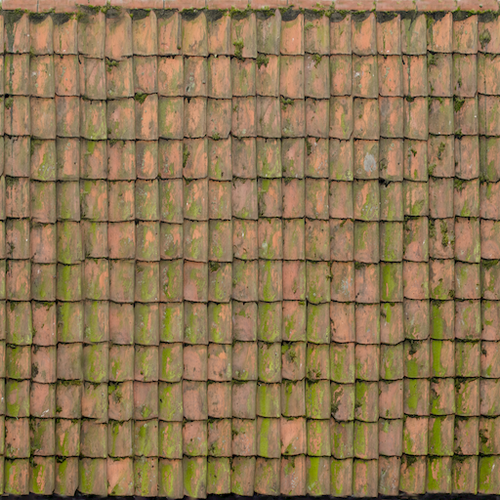} &
            \includegraphics[width=\imgW]{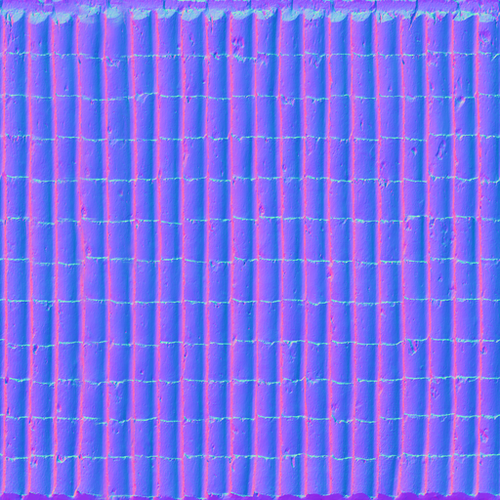} &
            \includegraphics[width=\imgW]{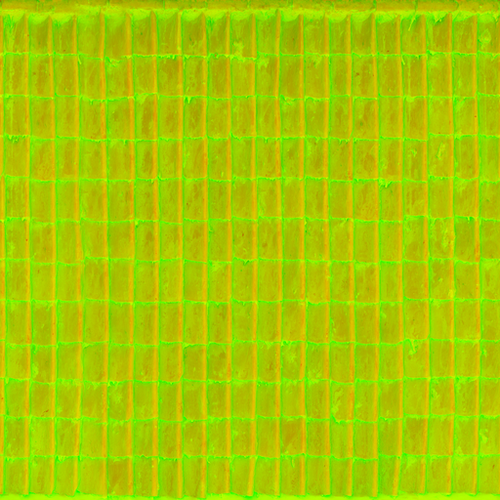} &
            \includegraphics[width=\imgW]{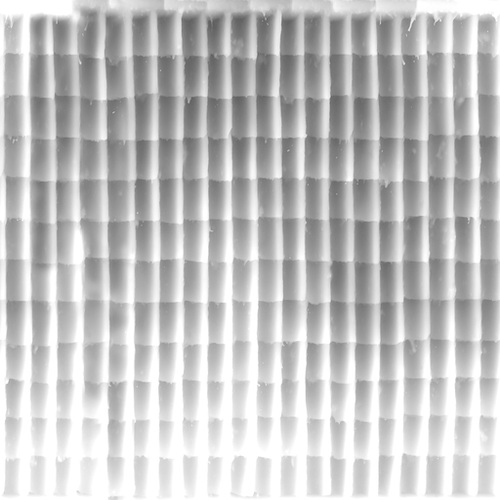}
            \\
            diffuse/albedo & normal map & ARM & displacement map
        \end{tabular}
        }
    \caption{An example \textit{texture set} consisting of
    a diffuse map, normal map, an ARM (ambient occlusion, roughness, metalness) texture, and
    a displacement map, for a ceramic roof material. Our approach compresses these textures together.
    Textures retrieved from \url{https://polyhaven.com/}.
    }
    \Description{TODO}
    \label{fig_textureset_example}
\end{figure}

\subsection{Neural Rendering and Materials}
Neural rendering~\cite{tewari2022advances} is a new field that has emerged recently and
includes approaches that leverage neural methods inside traditional rasterization or ray tracing based renderers, which is particularly
relevant to our work.
Thies et al.~\cite{thies2019deferred} proposed a method for higher quality image synthesis from low quality 3D content, by storing learned neural features in a texture, which are then sampled and rasterized into an off-screen buffer. The final images are then produced using a jointly-optimized neural renderer.

The idea of using neural latent grids to store spatially-varying appearance has been adopted by computer vision research, as well as computer graphics, where it has been used to represent complex materials and their mipmap chains~\cite{kuznetsov2021neumip}.
Most of these works focus on modeling materials and their appearance, using representations that can be significantly larger than traditional 8-bit textures.
While our study focuses on practicality and cost-efficiency in traditional rendering, it is important to note that our algorithm is also applicable to neural rendering, where it could greatly reduce memory consumption.

Neural networks have emerged as a popular alternative to discrete grids for signal representation. The predominant architectures are coordinate networks~\cite{mildenhall2021nerf}, which offer a fully differentiable and smooth representation, advantageous for 3D computer vision reconstruction tasks. Coordinate networks frequently employ \textit{positional encoding}, a concept originating from language modeling literature~\cite{vaswani2017attention}.
Instead of passing the input coordinates $\mathbf{p}$ directly to the MLP, this method encodes it as a vector of $\sin(2^h \pi \mathbf{p})$ and $\cos(2^h \pi \mathbf{p})$ terms, where $h$ represents an octave.
Fourier encoding has been shown to overcome the low-frequency bias of MLPs~\cite{tancik2020fourier}. 
For improved computational efficiency, trigonometric functions can be replaced by triangle waves~\cite{muller2021real}.
As an alternative to positional encoding, trigonometric~\cite{sitzmann2020implicit}, Gaussian~\cite{chng2022gaussian}, or wavelet~\cite{saragadam2023wire} activation functions can be used to increase the bandwidth of each layer.
This property can be used to band limit network parts and interactively stream only lower frequencies of the encoded content~\cite{lindell2022bacon}.

Storage size is often overlooked in neural representation literature, with typical coordinate networks requiring more storage to represent discrete 2D signals than their uncompressed form.
For instance, on the task of image fitting, the original work on positional-encoded coordinate networks~\cite{tancik2020fourier} uses a network with 
327K parameters to represent 197K scalar values (a $256 \times 256$ RGB image).
Similarly, work on periodic activation functions~\cite{sitzmann2020implicit}
uses 329K parameters to represent 786K scalar values (a $512 \times 512$ RGB image). For this representation size, both approaches report only a PSNR under 30 dB.
Improving the storage efficiency of these coordinate-based networks was a key motivation for subsequent work that used grid-based neural representations compressed using vector quantization~\cite{takikawa2022variable} or hash tables~\cite{muller2022instant}.

Finally, previous work on neural radiance caches~\cite{muller2021real} has shown that it is possible to efficiently embed small neural networks inside a renderer, enabling training and inference in real time, orders of magnitude faster than traditional deep learning frameworks. We build on this work and optimize it further.

\begin{figure}[t]
\adjustbox{width=0.9\columnwidth,trim=0cm 0cm 0cm 0cm,clip}
{
\input{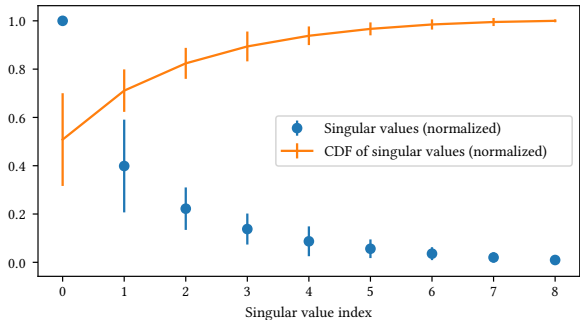}
}
\caption{Singular value distribution representing cross-channel correlation in 20 texture sets selected from diverse materials. The sharp falloff in the singular values indicates a high degree of correlation across channels.}
\label{fig:fig_singular}
\vspace{-1em}
\end{figure}

\section{Motivation}\label{motivation}
Lossy image compression techniques typically exploit both spatial and cross-channel correlations and, subsequently, quantize or eliminate weakly correlated information.
For example, the BC1 compression format maps all RGB color triplets in a $4\times4$ texel tile to a single line in RGB space, assuming perfect correlation between all channels.
Other BCx formats use similar assumptions but relax some constraints, such as the allowed number of lines inside a block.
Block compression formats, however, can only compress textures with up to four channels, while
modern renderers typically use several material properties, including diffuse color, normals maps, height maps, ambient occlusion, glossiness, roughness, and other BRDF information. 
These properties are typically stored as multiple textures within a group, which we refer to as a \textit{texture set} (Figure~\ref{fig_textureset_example}). As seen in Figure~\ref{fig:fig_singular}, there is significant correlation across the channels of different textures in a texture set. 
This can be attributed to both the physical properties of real-world materials (specularity and albedo are inversely correlated), geometric properties (displacement, normal maps, edges), as well as the material authoring process, where an artist may layer, mask, and combine multiple channels together~\cite{neubelt2013crafting}.

Earlier work~\cite{wronski2020,wronski2021} has noted this correlation, applying it to dimensionality reduction of material inputs. 
Besides correlations across pixels and channels, Zontak et al.~\shortcite{zontak2011internal} have also noted redundancies across multiple scales.
In this paper, we derive a neural compression scheme that builds on these observations and exploits redundancies
spatially, across mip levels, and between all channels of a texture set.

\begin{figure*}[h] 
	\centering
	\includegraphics[width=\textwidth, trim=0 9.6cm 6cm 0, clip]{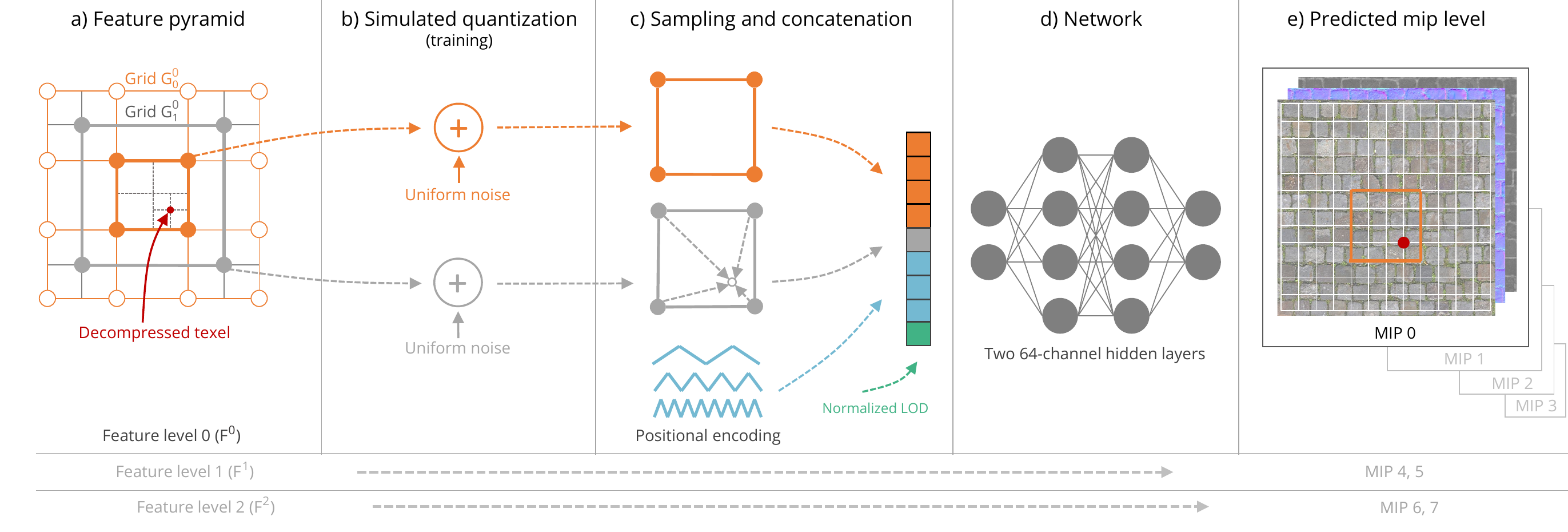}
	\caption{Overview of our method.
 a) Our compressed representation comprises multiple feature levels, each having two feature grids; a high resolution grid $G_0$ and a low resolution grid $G_1$ (Section~\ref{pyramid}). The solid circles represent the grid cells accessed for a target texel (in red).
 b) During training, we simulate quantization through addition of noise and clipping (Section ~\ref{quantization}).
 c) During inference and training, we sample the four neighboring feature vectors (orange circles) from the grid $G_0$ (Section~\ref{feature_interpolation}) and bilinearly interpolate features from $G_1$ (hollow gray circle), concatenating them with local positional encoding (Section ~\ref{posenc}) and a normalized level-of-detail (LOD) value for the target mip level.
 d) Finally, we use a neural network (Section~\ref{mlp}) to decode the mip level (e).
 }
	\label{fig:ntc}
\end{figure*}

\section{Neural Material Texture Compression} \label{method}

We represent the texture set as a tensor with dimensions $w\times h \times c$ and our model compresses the tensor without making any  assumptions about the channel count or the specific semantics of each channel. For example, the normals or diffuse albedo could be mapped to any channels without affecting compression. 
This is possible because we learn the compressed representation for each material individually, effectively specializing it for its unique semantics. 
The only assumption we make is that each texture in a texture set has the same width and height. 
Some materials can have BRDF properties not present in other ones, for instance, subsurface scattering color or thickness.
While an alternative approach using a pre-trained global encoder could potentially achieve faster compression, it would also require imposing globally pre-assigned semantics for each channel, which can be impractical for a large set of diverse materials. 

Figure~\ref{fig:ntc} illustrates the decoding process, progressing from a compressed representation, on the left, to a decompressed texel, on the right. Our compressed representation is a pyramid of quantized features levels, which are typically at a lower resolution compared to the reference texture. To achieve decompression of a single texel, feature vectors are sampled from a feature level and subsequently decoded to generate all channels within the texture set.
To facilitate greater feature decorrelation, the decoder is modeled as a non-linear transform~\cite{balle2020}, utilizing a multilayer perceptron (MLP) as a universal approximator~\cite{hornik1989multilayer}. This MLP is shared across all the mipmap (mip) levels, which enables joint learning of the compressed representation and the MLP's weights, using an autodecoder framework~\cite{park2019_cvpr}. Specifically, the compressed representation is directly optimized through quantization-aware training and backpropagation through the decoder, as opposed to using an encoder.

The following sections describe each stage of decompression in detail. 
Later, in Section~\ref{sec:results}, we show how those assumptions hold over a diverse set of materials and textures from different datasets, different formats, and using different material semantics.

\subsection{Feature Pyramid}
\label{pyramid}

As shown in Figure~\ref{fig:ntc}~(a), our compressed representation is a pyramid of multiple \emph{feature levels} $F^j$, with each level, $j$, comprising a pair of 2D \emph{grids}, $G^j_0$ and $G^j_1$. 
The grids' cells store ~\emph{feature vectors} of quantized latent values, which are utilized to predict multiple mip levels. This sharing of features across two or more mip levels lowers the storage cost of a traditional mipmap chain from $33\%$ to $\sim\!6.7\%$ or less.
Furthermore, within a feature level, grid $G_0$ is at a higher resolution, which helps preserve high-frequency details, while $G_1$ is at a lower resolution, improving the reconstruction of low- frequency content, such as smooth gradients.

Table~\ref{tab_features} illustrates the feature levels and grid resolutions for a $1024\times1024$ texture set. The resolution of the grids is significantly lower than the texture resolution, resulting in a highly compressed representation of the entire mip chain. Typically, a feature level represents two mip levels, with some exceptions; the first feature level must represent all higher resolution mips~(levels 0 to 3), and the last feature level represents the bottom three mip levels, as it cannot be further downsampled.

\begin{table}[b]
\centering
\caption{Compressed representation of a mip chain through feature levels and low resolution grids for a $1024\times1024$ texture set.}
\vspace*{\tabvspace}
\resizebox{\columnwidth}{!}{%
\begin{tabular}{@{}c|c|c|c@{}}
\toprule
Feature level $F^j$ & $G^j_0$ grid resolution & $G^j_1$ grid resolution & Predicted mip levels \\ \midrule
0 & 256$\times$256 & 128$\times$128 & 0,1,2,3 \\
1 & 64$\times$64 & 32$\times$32 & 4,5 \\
2 & 16$\times$16 & 8$\times$8 & 6,7 \\
3 & 4$\times$4 & 2$\times$2 & 8,9,10 \\ \bottomrule
\end{tabular}%
}
\label{tab_features}
\end{table}

\subsection{Simulated Quantization}\label{quantization}

Since we do not use entropy coding, we enforce a fixed quantization rate for all latent values in a feature grid and only optimize for image distortion. 
We simulate quantization errors along the lines of previous neural image compression techniques~\cite{balle2017} by adding uniform noise in the range $\left(-\frac{Q_k}{2}, \frac{Q_k}{2}\right)$ to the features, where $Q_k$ is the range of a quantization bin
on grid $k$. 
To limit the number of quantization levels, we clamp features to the quantization range after updating them in the backward pass. This ensures that both gradient computations and feature updates are w.r.t.\ values strictly within the quantization range. 
We observed that this approach produces better results than clamping features in the forward pass, where the learned features can drift outside the desired quantization range.

For each feature grid $G^j_k$, we use an asymmetric quantization range of $\left[-\frac{N_k-1}{2}Q_k, \frac{N_k}{2}Q_k\right]$, where $N_k = 2^{B_k}$ is the desired number of quantization levels. This quantizes a zero value with no errror by aligning it with the center of a quantization
bin~\cite{jacob2018cvpr}. In turn, this produces better results especially when we quantize to four levels or less. We set $Q_k$ to $\frac{1}{N_k}$ and therefore $N_k$ is the only value provided during training.
Toward the end of the training process, we stop adding noise to simulate quantization and explicitly quantize the feature values. The feature values are frozen for the rest of the training. Then, we continue to optimize the network weights for 5\% more steps, adapting them to the discrete-valued grids. 
We also include a comparison between scalar quantization and vector quantization~\cite{oord2017} in our supplementary material (Appendix~\ref{sec:vq_comparison}).

\subsection{Sampling and Concatenation}
\label{sec_sampling_concatenation}

In this section, we describe the first stage of decompression, which samples the grids of a feature level and prepares the input to the MLP, as shown in Figure~\ref{fig:ntc}~(c).
In this stage, we first select a feature level based on the desired level of detail (LOD) (Table~\ref{tab_features}), and then resample both the grids in the feature level to the target resolution. 
In the next section, we describe how grids are resampled by interpolating the features at the target texel location. Following this, we describe our positional encoding scheme that aids in interpolation and preserving high-frequency details. 

\subsubsection{Feature Interpolation}\label{feature_interpolation}

Features may be upsampled or downsampled depending on the feature level and the target LOD. However, upsampling the first feature level $F^0$ alone presents the main challenge for reconstruction quality, as it is typically at a much lower resolution than the input texture. To a large extent, we rely on  the lower resolution of the grids for compression.

\vfill\null
To achieve real-time decompression performance, we balance complexity against reconstruction quality by using two different approaches for resampling the grids. 
We use a \emph{learned interpolation} approach for the higher resolution grid $G_0$ and bilinear interpolation for the lower resolution grid $G_1$.
In the case of learned interpolation, we concatenate four neighboring feature vectors and rely on phase information from the positional encoding (Section~\ref{posenc}) to reconstruct
high-frequency details. Concatenation, as opposed to summation of weighted features, allows the following MLP layers to combine features differently depending on the texel location.
However, the learned interpolation also increases the cost of the input layer of the network. The bilinear interpolation of the low resolution grid was chosen to limit this. We observed that the smooth output of bilinear interpolation can compliment learned interpolation well by suppressing banding artifacts resulting from heavily quantized features.

\subsubsection{Tiled Positional Encoding}\label{posenc}

To improve the fidelity of high-frequency details, we condition our decoder on \textit{positional encoding}~\cite{vaswani2017attention,mildenhall2021nerf}.
We use a more computationally efficient variant of the encoding~\cite{muller2021real}, which is based on triangular waves, and observe no quality loss.

\begin{figure}[t]
\includegraphics[width=\linewidth]{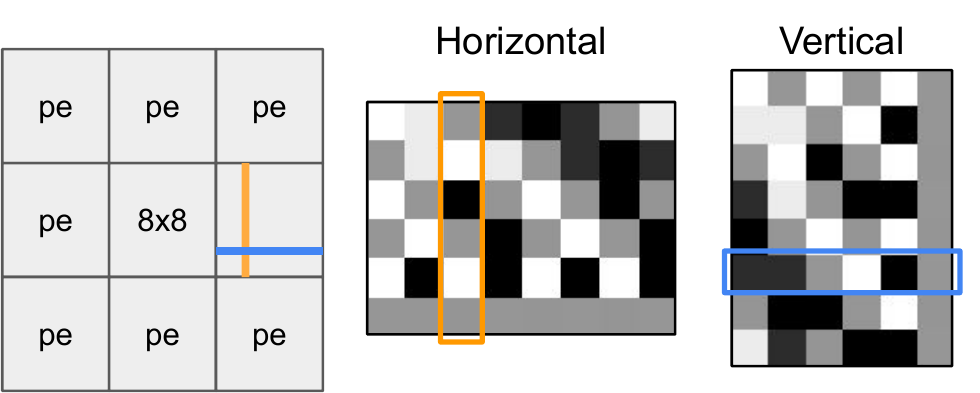}
\caption{Positional encoding tiles of $8\times8$ texels. A single texel is represented by 6+6 scalars, each encoding the horizontal and vertical texel position inside the tile. The last value is constant in both the horizontal and vertical encoding.}
\label{fig:fig_posenc}
\end{figure}

Our architecture is not fully coordinate-based since we also use features stored in low-resolution grids.
Therefore, any low-frequency information can be directly represented by the features and we only need positional encoding to represent frequencies higher than the Nyquist limit of the grids. The number of octaves for the encoding is $\log_2 8$, as 8 is the maximum upsampling factor we encounter, i.e., when upsampling grid $G_1$ to a target LOD of 0. Consequently, the encoding is a tiled pattern that repeats every $8\times8$ texels, as shown in Figure~\ref{fig:fig_posenc}.

\begin{figure}[th]
    \centering
    \newcommand{\imgW}{0.15\textwidth}
    \setlength{\tabcolsep}{1.0pt}%
    \begin{tabular}{ccc}
        \includegraphics[angle=90,width=\imgW]{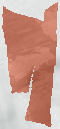} &
        \includegraphics[angle=90,width=\imgW]{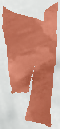} &
        \includegraphics[angle=90,width=\imgW]{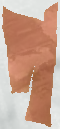}
    \end{tabular}
  \caption{With a low-bitrate model, the choice of loss function
can improve color fidelity or preservation of high-frequency details but typically not both. \textbf{Left:} Original.
\textbf{Middle:} L2. \textbf{Right:} Linear combination of L2 and $1-\textrm{SSIM}$. Textures
retrieved from \url{https://ambientcg.com}.}
  \label{fig:fig_loss_comparison}
\end{figure}
\vspace{2em}

\subsection{Network}    %
\label{mlp}

Our network is a simple multi-layer perceptron with two hidden layers, each of size 64 channels. 
The size of our input is given by $4C_0 + C_1 + 12 + 1$, where $C_k$ is the size of the feature vector in grid $G_k$. Note that we use 4$\times$ more features from grid $G_0$ for learned interpolation, 12 values of positional encoding and a LOD value.

We do not use any activation function on the output of the last layer.  
We experimeted with several different activation functions for the three remaining layers and observed best results with GELU~\cite{hendrycks2016gelu}.
To reduce the computation overhead of GELU functions, we derived an approximation denoted ``hardGELU'', which is similar to hard Swish~\cite{mobilenetv3}. Our variant is given by
\[ \textrm{hardGELU}(x) =
  \begin{cases}
    0,  & \quad \text{if } x < -\frac{3}{2},  \\
    x,  & \quad \text{if } x > \frac{3}{2}, \\
    \frac{x}{3}(x + \frac{3}{2}),   & \quad \text{otherwise.} \\
  \end{cases}
\]

\subsection{Optimization Procedure and Loss Function}\label{optimization_and_loss}
We jointly optimize the feature pyramid and the decoder, using gradient descent with the ADAM~\cite{kingma2014adam} optimizer. Unless stated otherwise, our model is trained for 250k iterations.
Our method can use and minimize an arbitrary image loss function.

To optimize our compressed representation, we explored several different loss functions,
including SSIM~\cite{Wang2004a}, a version of VGG loss that supports texture sets~\cite{Chambon2021}, adversarial
as well as L1 and L2 losses,
and combinations thereof.
In general, we found that 
the loss function presented a 
compromise between maintaining
color fidelity and preserving high-frequency details
-- though we were unable to find a loss
function that did not show weaknesses
in one or the other.
Figure~\ref{fig:fig_loss_comparison}
illustrates this behavior, where using only L2
results in loss of high-frequency detail, while adding SSIM
improves on this but discolors the image.
The choice of objective function can thus be adapted based on
the use case and
when the application only
requires one of the two quality properties
to be preserved. We found the L2 loss to be a reasonable compromise.
As it also trains robustly and is the simplest and computationally fastest choice, we use it throughout this paper.

We hypothesize that the observed behavior
is a consequence of information theoretical
limitations, i.e., that we cannot preserve
both high-frequency detail and color fidelity
at this low bitrate. Further investigation
of this hypothesis is left for future work.
In addition, we conducted initial
experiments to explore potential benefits of
using different specialized objective functions
for different texture types. While these experiments
did not indicate advantages of such an approach,
we believe it to be an interesting area of future research.

\section{Implementation}
As outlined in in Section~\ref{method}, we decompress textures at a given texel by sampling the corresponding latent values from a feature pyramid and decoding them using a small MLP network. 
Our compressed representation, as mentioned previously, is trained specifically for each texture set.
Specializing the compressed representation for each material %
allows for using smaller decoder networks
, resulting in fast optimization (compression) and real-time decompression. 

\begin{figure}[t]
\includegraphics[width=0.9\columnwidth, trim=1.5cm 7cm 7.2cm 1.4cm, clip]{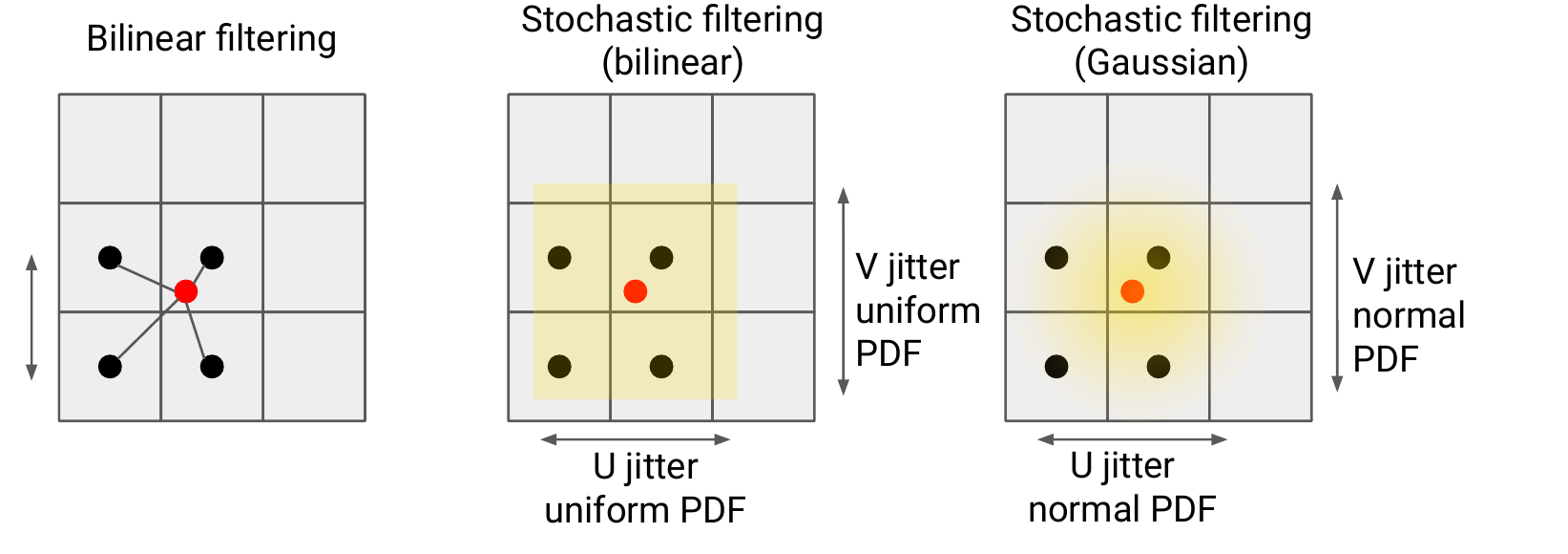}
\caption{Bilinear and stochastic filtering.}
\label{fig:fig_stochastic_filter}
\vspace{-1em}
\end{figure}

\subsection{Compression}
\label{sec:compression}

Similar to the approach used by M\"uller~et~al. for training autodecoders~\shortcite{muller2022instant}, we achieve practical compression speeds by using half-precision tensor core operations in a custom optimization program written in CUDA.
We fuse all of the network layers in a single kernel, together with feature grids sampling, loss computations, and the entire backward pass. This allows us to store all network activations in registers, thus eliminating writes to shared or off-chip memory for intermediate data. 

We process batches of eight randomly sampled $256\times 256$ texel crops, selected from the same level of detail. 
For each batch, we randomly choose a level of detail proportionally to the mip level's area by sampling from an exponential distribution:
$\text{LOD} = \lfloor -\log_{4} X \rfloor$, where $X \sim U(0, 1).$
To mitigate undersampling of low-resolution mip levels, $5\%$ of the batches sample their LOD from a uniform distribution defined over entire range of the mip chain.
We use a high initial learning rate of $0.01$ for the latent grids, and a lower value of $0.005$ for the network weights and apply cosine annealing ~\cite{sgdr2017}, lowering the learning rate to 0 at the end of training.  

\subsection{Decompression}

Inlining the network with the material shader presents a few challenges as matrix-multiplication hardware such as tensor cores operate in a SIMD-cooperative manner, where the matrix storage is interleaved across the SIMD lanes~\cite{matrixfragment, yan2020}.
Typically, network inputs are copied into a matrix by writing them to group-shared memory and then loading them into registers using specialized matrix load intrinsics. 
However, access to shared memory is not available inside ray tracing shaders. 
Therefore, we interleave the network inputs \emph{in-registers} using SIMD-wide shuffle intrinsics.

We used the Slang shading language~\cite{Slang2018} to implement our fused shader along with a modified Direct3D~\shortcite{direct3d} compiler to generate NVVM~\cite{Nvvm} calls for matrix operations and shuffle intrinsics, which are currently not supported by Direct3D. These intrinsics are instead directly processed by the GPU driver. 
Although our implementation is based on Direct3D, it can be reproduced in Vulkan~\shortcite{vulkan} without any compiler modifications, where  accelerated matrix operations and SIMD-wide shuffles are supported through public vendor extensions.
The NV\textunderscore cooperative\textunderscore matrix extension~\shortcite{nvcoopmatrix} provides access to matrix elements assigned to each SIMD lane. The mapping of these per-lane elements to the rows and columns of a matrix for NVIDIA tensor cores is described in the PTX ISA~\shortcite{matrixfragment}. The KHR\textunderscore shader\textunderscore subgroup extension~\shortcite{glsubgroup} enables shuffling of values across SIMD lanes in order to assign user variables to the rows and columns of the matrix and vice versa. These extensions are not restricted to any shader types, including ray tracing shaders.

\subsubsection{SIMD Divergence} 
In this work, we have only evaluated performance for scenes with a single compressed texture-set. 
However, SIMD divergence presents a challenge as matrix acceleration requires uniform network weights across all SIMD lanes. This cannot be guaranteed since we use a separately trained network for each material texture-set. 
For example,  rays corresponding to different SIMD lanes may intersect different materials. 

In such scenarios, matrix acceleration can be enabled by iterating the network evaluation over all unique texture-sets in a SIMD group. 
The pseudocode in Appendix~\ref{handling_divergence} describes divergence handling. SIMD divergence can significantly impact performance and techniques like SER~\shortcite{Ser} and TSU~\shortcite{Tsu} might be needed to improve SIMD occupancy. A programming model and compiler for inline networks that abstracts away the complexity of divergence handling remains an interesting problem and we leave this for future work.

\begin{figure}[t]
\includegraphics[keepaspectratio, width=\linewidth]{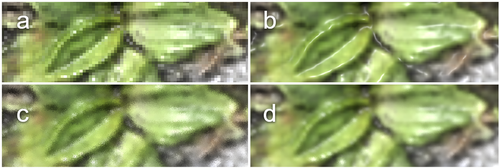}
\caption{Filtering across the boundaries of a highly specular material. 
\textbf{a)} Nearest-neighbor filtering. \textbf{b)} Trilinear filtering. \textbf{c)} Single frame of stochastic filtering. \textbf{d)} Resolved stochastic temporal filtering. Trilinear texture filtering causes specular lighting artifacts as the high specularity interpolates outside the glossy material.
Textures retrieved from \url{https://ambientcg.com}.}
\label{fig:fig_filtering_comparison}
\end{figure}

\begin{figure*}[t]
    \adjustbox{width=\linewidth,trim=0cm 0.2cm 0cm 0cm,clip}{\input{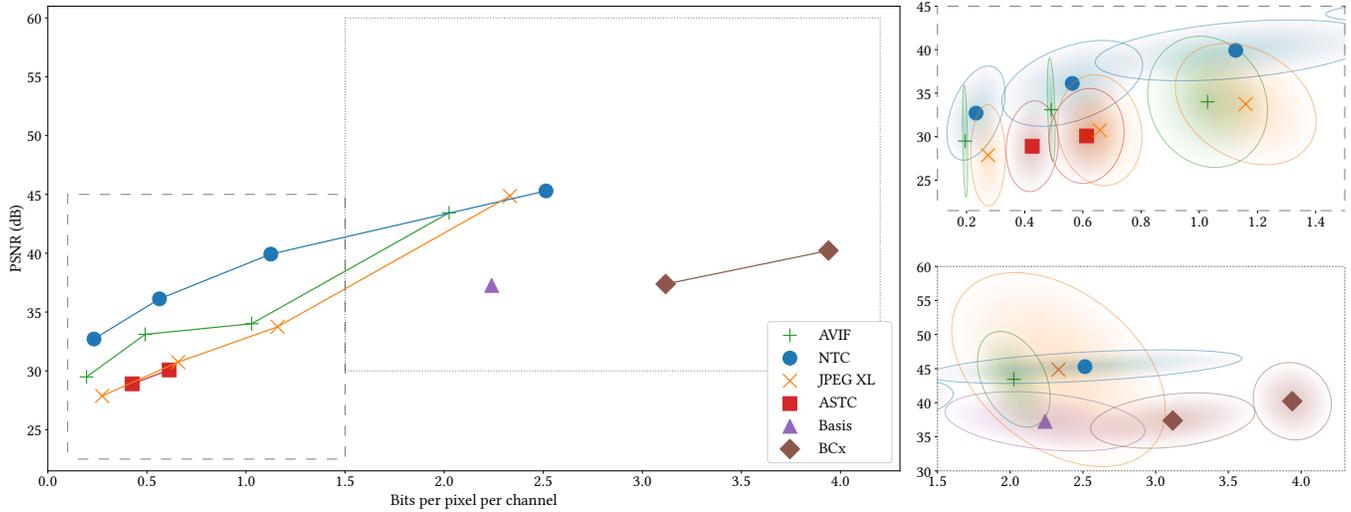}}
    \caption{Quantitative results. Vertical axis: PSNR scores of different methods. The horizontal axis is bits per pixel per channel storage cost.
    The size of each ellipse corresponds to the variation in PSNR and compressed material size due to different texture channel counts present in the data set. Table~\ref{fig:table_methods_comparison} contains
    the explicit numbers for the markers in this figure,
    in addition to corresponding SSIM and LPIPS values.}
    \label{fig:fig_quantitative2}
\end{figure*}

\subsection{Filtering}
\label{sec:filtering_implemetation}

Our method supports mipmapping for discrete levels of minification (Section~\ref{pyramid}), similar to BCx compression methods.
For the best quality and compression ratios, our compression approach relies on overfitting the network only at the discrete texel locations in the original texture. However, this does not guarantee smooth reconstruction in between these discrete texel locations and mipmap levels.
Therefore, we cannot rely on hardware acceleration for trilinear filtering and we implement it in software on the GPU.
The software implementation decompresses and combines four texels for bilinear filtering, and eight texels for trilinear filtering, significantly increasing decompression cost.

In order to decouple the decompression cost from filtering, we propose a simple alternative to trilinear filtering based on stochastic sampling~\cite{stochastic_convolution, Ernst2006, Hofmann2021}, which we call~\emph{stochastic filtering}.
We add random noise to the $(u,v)$ position, followed by nearest neighbor sampling. We can achieve different types of texture filtering by changing the distribution  of the noise, as shown in Figure~\ref{fig:fig_stochastic_filter}. 
For example, a uniform distribution in the range $(-0.5, 0.5)$ of one texel produces bilinear filtering, and a normal distribution produces Gaussian filtering.
In addition to jittering the $(u,v)$ coordinates, we also jitter the LOD to enable a smooth transition between mip levels.

Stochastic filtering typically increases the amount of noise in the rendered image, but we observed that modern post-process reconstruction techniques~\cite{dlss2022,fsr2022,xess2022} can effectively suppress this noise. 
Figure~\ref{fig:fig_filtering_comparison} shows a comparison of trilinear filtering and stochastic filtering with reconstruction using DLSS~\cite{dlss2022}.
We note that, while traditional texture filtering filters the input material properties, stochastic filtering filters the shading output, which produces more accurate results, as shown in Figure~\ref{fig:fig_filtering_comparison}.

\section{Results}
\label{sec:results}

Our neural compression method for material textures (NTC)
is flexible, allowing for adjustments in feature grids resolution, 
channel counts, and bit depths to optimize the trade-off between compression quality, 
storage, and inference performance.
We use the four compression profiles listed in Table~\ref{table:compressionprofiles}, each targeting a different number of bits per-pixel per-channel (BPPC), for our comparisons. 
Table~\ref{table:compressionprofiles} also shows the total storage cost including network weights and feature levels. 
The cost of the network weights is roughly constant across all profiles, and does not change with the texture resolution. Please refer to Appendix~\ref{sec:storage_cost} in the supplementary material for
more details of the NTC storage cost.

\subsection{Evaluation Data Set}
\label{sec:evaluation_dataset}
To evaluate different compression techniques, we selected 20 diverse materials with texture sets varying in content, frequency characteristics, resolution, and channel counts.
The content includes natural textures, human-made objects, character textures, and a synthetic gradient, with attributes
such as high-frequency repeating patterns, noisy, and smooth. The texture sets have resolutions varying from $2048 \times 2048$ to $8192 \times 8192$ texels and channel counts ranging from 3 to 12.
More details about the evaluation data set can be found in our supplementary material (Appendix~\ref{sec_texture_set_selection}).

\begin{table}[b]
\centering
\caption{NTC profiles with different bits per-pixel per-channel (BPPC) and total storage costs calculated using a $4096\times4096\times9$ texture set as reference
}
\vspace*{\tabvspace}
\resizebox{\columnwidth}{!}{%
\begin{tabular}{@{}l|c|c|c|c@{}}
\toprule
BPPC Profile & $G^0_0$ grid resolution & $G^j_0$ grid channels & $G^j_1$ grid channels & Total (MB) \\ \midrule
NTC 0.2 & 1024$\times$1024 & 8$\times$2b & 12$\times$4b & 3.52 \\
NTC 0.5 & 1024$\times$1024 & 12$\times$4b & 20$\times$4b & 8.53 \\
NTC 1.0 & 2048$\times$2048 & 12$\times$2b & 10$\times$4b & 17.03 \\ %
NTC 2.25 & 2048$\times$2048 & 16$\times$4b & 12$\times$4b & 38.03 \\ \bottomrule
\end{tabular}%
}
\label{table:compressionprofiles}
\end{table}

\begin{table*}[t]
\caption{Average PSNR, $1-\textrm{SSIM}$,
and LPIPS values over the evaluation data set for the
methods used in our comparison. The methods are
grouped based on their storage requirements.
Algorithms marked in gray do not support random access.
``BC~M.'' is short for ``BC medium'' and ``BC~H.'' is for ``BC high.''
}
\vspace*{\tabvspace}
\resizebox{\textwidth}{!}{%
\begin{tabular}{@{}r|gcg|cgccg|gcg|gcgcc|c@{}}
\toprule
\multicolumn{1}{l|}{} &
\multicolumn{3}{c|}{Low ($\sim$ 0.2 BPPC)} &
\multicolumn{5}{c|}{Medium-low ($\sim$ 0.5 BPPC)} &
\multicolumn{3}{c|}{Medium ($\sim$ 1.0 BPPC)} &
\multicolumn{5}{c|}{Medium-high ($\sim$ 1.5 - 3.0 BPPC)} &
\multicolumn{1}{c}{High ($\sim$ 4.0 BPPC)} \\ \midrule
& AVIF & NTC & JPEG XL & ASTC $12\times 12$ & AVIF & NTC & ASTC $10\times 10$ & JPEG XL & AVIF & NTC & JPEG XL & AVIF & Basis & JPEG XL & NTC & BC M. & BC H. \\ \midrule
BPPC (mean) & 0.20 & 0.23 & 0.27 & 0.43 & 0.49 & 0.56 & 0.61 & 0.66 & 1.03 & 1.13 & 1.16 & 2.02 & 2.24 & 2.33 & 2.51 & 3.12 & 3.94 \\ \midrule
PSNR (\textuparrow) & 29.49 &  \textbf{32.71} &  27.87 &  28.90 &  33.10 &  \textbf{36.12} &  30.08 &  30.74 &  34.01 &  \textbf{39.92} &  33.74 &  43.44 &  37.27 &  44.86 &  \textbf{45.30} &  37.38 &  40.23 \\
1 - SSIM (\textdownarrow) &  0.1138 &  \textbf{0.0633} &  0.1377 &  0.1034 &  0.0586 &  \textbf{0.0375} &  0.0788 &  0.0778 &  0.0380 &  \textbf{0.0183} &  0.0451 &  0.0068 &  0.0219 &  \textbf{0.0030} &  0.0076 &  0.0211 &  0.0099 \\
LPIPS (\textdownarrow) &  0.1051 &  \textbf{0.0660} &  0.1001 &  0.0722 &  0.0302 &  0.0364 &  0.0460 &  \textbf{0.0272} &  0.0108 &  0.0176 &  \textbf{0.0087} &  0.0013 &  0.0114 &  \textbf{0.0003} &  0.0057 &  0.0133 &  0.0024  \\ \bottomrule
\end{tabular}%
}
\label{fig:table_methods_comparison}
\vspace{-1em}
\end{table*}

\begin{figure*}[ht]
\includegraphics[width=\linewidth, trim=0cm .3cm 0cm 0cm, clip]{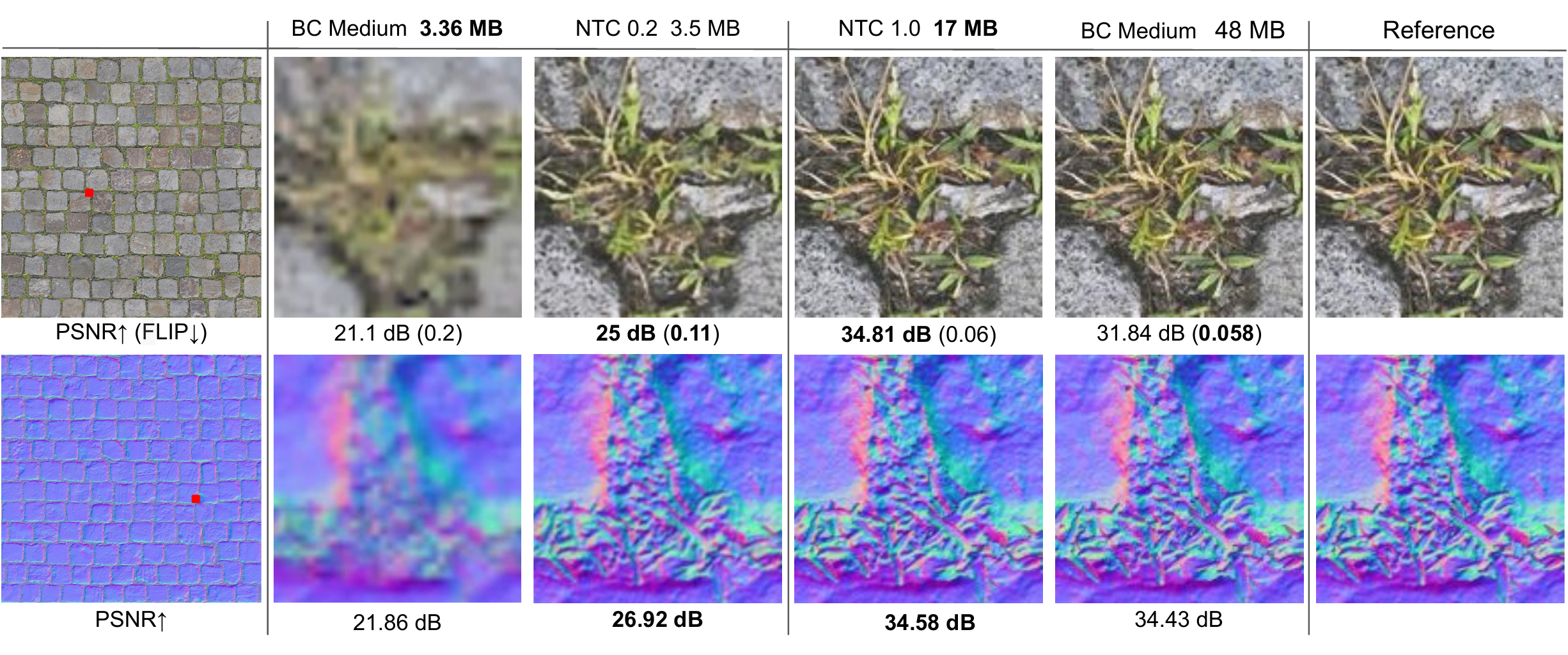}
\vspace{-2em}
\caption{Iso-storage and iso-quality comparison showing PSNR and \FLIP scores for the diffuse and normal map textures in the \emph{Paving Stones} texture set,
retrieved from \url{https://ambientcg.com}. For iso-storage comparisons, we use two higher mip levels for the BCx textures to match the storage size of NTC.}
\centering
\label{fig:iso_storage_quality}
\end{figure*}

\subsection{Compared Methods}

Our method can replace
GPU texture compression techniques, such as BC~\cite{BCCompression2020} and ASTC~\cite{Nystad2012}. %
It is a common industry practice to use different BC variants for different material texture types~\cite{graphicsstudies20}, but there is no single standard.
As such, we propose two compression profiles for the evaluation of BC, %
namely ``BC medium'' and ``BC high.'' The BC medium profile uses BC1 for diffuse and other packed multi-channel textures, BC7 for normals, and BC4 for any remaining single-channel textures. The BC high profile, on the other hand, uses BC7 for
three-channel textures and BC4 for one-channel textures.

Our method is not directly comparable with compression formats using entropy encoding, as NTC is designed to support real-time random access. 
However, to provide a frame of reference for storage and quality at lower bitrates,
we also evaluate methods using entropy encoding, namely, the texture meta-compression algorithm
Basis Universal~\cite{Hurlburt2022}
and the recently standardized, high-quality image compression formats AVIF and JPEG XL.
We have not included recent neural image compression techniques in our analysis as they do not produce significantly better quantitative results compared to traditional methods on metrics like MSE, despite typically offering better perceptual characteristics~\cite{cheng_2020_cvpr}.
In addition, they do not support random access, which is a key property for GPU texture accesses.
Details of the compression settings used for our evaluation are included in our supplementary material (Appendix~\ref{supplement_compression_artifacts}).

\subsection{Quantitative Results}
\label{sec:quantitative_results}

Figure~\ref{fig:fig_quantitative2} presents an overview of our compression results, showing PSNR values for different compression methods, across a range of BPPC rates as well as the variations across texture sets.
The PSNR values are computed over the entire texture set and all mip levels down to resolution $4\times 4$
(see Appendices~\ref{sec:error_computation} and \ref{sec_texture_set_selection} in our supplementary material).
The plot shows that our method significantly outperforms GPU texture compression at high bitrates and surpasses advanced compression methods at lower rates of less than 1.5 BPPC.
We attribute these improvements to significant cross-channel and cross-mip-level correlations, not exploited in prior works.

Our compression method can significantly reduce texture sizes, which can be leveraged in different ways. For example, the NTC 0.5 profile can be used to achieve iso-quality results as the BC medium profile, at 1/5th of the storage cost. 
Alternatively, the NTC 0.2 profile can be used to store two additional higher mip levels at a PSNR that is slightly lower but still better quality than AVIF and JPEG XL. This enables significantly higher detail, as demonstrated in Figure~\ref{fig_teaser}.

Although our compression is not optimized for perceptual quality, we also include SSIM~\cite{Wang2004a} and LPIPS~\cite{Zhang2018} metrics in Table~\ref{fig:table_methods_comparison} for comparison. Even with these perceptual metrics, NTC provides better results than all other compression methods at low and medium-low rates, only trailing recent high-quality image compression techniques at higher rates. 
Perceptual quality of our method can be further improved by optimizing for perceptual metrics. 

The quantitative results with our compression technique are also consistent across different mip levels and different material textures. We include per-mip and per-texture-type PSNR results in our supplementary material (Appendix~\ref{app:quantitative}).

\subsection{Qualitative Results} 
\label{sec:qualitative_results}
Previous work has noted that the PSNR metric is not sufficient for image quality comparison and, furthermore, that \textit{objective} distortion metrics are at an inherent trade-off with perceptual quality~\shortcite{blau2018perception}.
Unfortunately, there is no single metric that would perfectly correlate with human preferences, which might vary for different applications.
We use qualitative, human analysis to evaluate image quality in addition to the PSNR metric. We also include \FLIP~\shortcite{andersson2020flip} values, which are better aligned with perceptual quality.

\subsubsection{Texture Quality}

A key motivation for our work was to determine the extent to which we could preserve image quality while reducing the storage. 
Figure~\ref{fig:iso_storage_quality} provides an overview of this by comparing the BC medium profile with NTC at different rates. 
We present an approximately \emph{iso-quality} comparison using the medium rate NTC 1.0 profile and \emph{iso-storage} comparison using the low rate using NTC 0.2 profile.
To evaluate BC medium at a low BPPC rate, we excluded the two largest mip levels to achieve a comparable storage size to NTC 0.2.
For both comparisons, we used the 4096$\times$4096 resolution \emph{Paving Stones} texture set, which is one of the most challenging in our evaluation. 

In the iso-quality comparison, we see that the BC medium profile consumes 48 MB of storage, while the NTC 1.0 profile exceeds its quality with just 1/3rd of the size.
We also see that the NTC 0.2 profile only consumes 3.5 MB of storage and has a significantly higher quality that appears closer to the reference than the BC medium profile at a comparable  size.

Figure~\ref{fig_big_comparison} presents a more extensive iso-storage comparison for the NTC 0.2. 
We selected the $8192\times 8192$, \emph{painted concrete} texture set for this comparison as it is close to the mean PSNR score reported in Table~\ref{fig:table_methods_comparison}.
Overall, our method produces results that are significantly better than BCx compression with the same storage.
We also observe better normal map quality compared to advanced image compression techniques like AVIF and JPEG XL, while the diffuse texture is slightly blurrier. %
We provide a more detailed qualitative analysis of this texture set in our supplementary material (Appendix~\ref{supplement_compression_artifacts}) and cover a set of failure cases in Section~\ref{sec:limiations}.

\begin{figure}[t]
\scalebox{0.6}{
\input{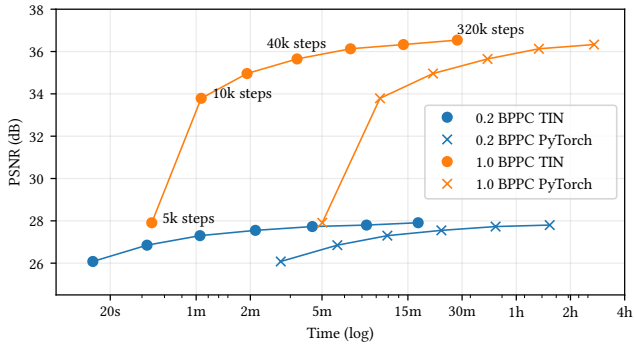}
}
\vspace{-1em}
\caption{
PSNR vs. training time for a 4096$\times$4096 texture set with 9 channels.
Our custom training implementation is an order of magnitude faster than PyTorch, 
compressing the texture set in a few minutes.
}
\label{fig:fig_training_time}
\vspace{-1em}
\end{figure}

\begin{table}[b]
\caption{Decompression performance for a 4k material texture set (\emph{Paving Stones}).
Performance is similar across all texture sets for a given profile.
}
\vspace*{\tabvspace}
\resizebox{\columnwidth}{!}{%
\begin{tabular}{@{}c|c|c|c|c@{}}
\toprule
\hspace{3mm} BC High \hspace{3mm} & \hspace{3mm} NTC 0.2 \hspace{3mm} & \hspace{3mm} NTC 0.5 \hspace{3mm} & \hspace{3mm} NTC 1.0\hspace{3mm}  & \hspace{3mm} NTC 2.25 \hspace{3mm} \\ \midrule
0.49 ms   & 1.15 ms   & 1.46 ms    & 1.33 ms    & 1.92 ms    \\ \bottomrule
\end{tabular}%
}
\label{decomp-perf}
\vspace*{\tabvspace}
\end{table}

\subsubsection{Rendered Quality}
Since our method is designed for compressing material textures used in rendering, we demonstrate the results with end-to-end rendered images in Figure~\ref{fig_teaser}.
The figure compares the NTC 0.2 profile, which uses 3.8~MB for the metal texture set, to ``BC high'' with two mip levels removed and still using 39\% more storage. As depicted in the insets, the quality of the NTC 0.2 profile is superior,  which is also indicated by the better PSNR and \FLIP numbers.
Our supplementary material includes additional examples, including a closed book and an open book with text, to further demonstrate the effectiveness of our method.

\subsubsection{Filtering Quality}
In the supplementary video, we showcase the quality of stochastic temporal filtering in motion. We use DLSS for high-quality spatiotemporal reconstruction~\cite{dlss2022} as described in Section~\ref{sec:filtering_implemetation}.
The use of temporal reconstruction techniques may exhibit flickering or ghosting in certain scenarios.
Our experiments reveal minor specular flickering, but no ghosting, under fast motion. 
A higher quality jitter sequence~\cite{wolfe22}, or reconstruction techniques optimized for stochastic filtering, could further improve quality.

\subsection{Performance}
In this section, we discuss compression performance as well as decompression performance in a simple renderer.

\subsubsection{Compression}

As described in Section~\ref{sec:compression}, we use a custom CUDA-implementation to optimize our compressed representation. 
Figure~\ref{fig:fig_training_time} shows our compression times for a single 4k material texture set with 9 channels, compared to a reference implementation in PyTorch~\cite{Paszke_2019_neurips}. Both implementations were evaluated on an NVIDIA RTX 4090 GPU for two different compression profiles.

Our custom implementation is approximately $10\times$ faster than PyTorch, which is crucial for achieving practical compression times.
We can generate a \textit{preview} quality result in just under one minute for both configurations, with a difference of less than 1.5~dB compared to the maximum length of optimization (320k steps) in the 0.2~BPPC case.
Moreover, for compression with the 1.0~BPPC profile, our implementation uses less than 2~GB of GPU memory, whereas PyTorch requires close to 18~GB, which is infeasible for many GPUs.

Traditional BCx compressors vary in speed, ranging from fractions of a second to tens of minutes to compress a single $4096\times 4096$ texture~\cite{arasp2020}, depending on quality settings.
The median compression time for BC7 textures is a few seconds, while it is a fraction of a second for BC1 textures. 
This makes our method approximately an order of magnitude slower than a median BC7 compressor, but still faster than the slowest compression profiles.

\subsubsection{Decompression} \label{sec:decompression}
We evaluate real-time performance of our method by rendering a full-screen quad at $3840\times2160$ resolution textured with the \emph{Paving Stone} set, which has 8 4k channels: diffuse albedo, normals, roughness, and ambient occlusion. The quad is lit by a directional light and shaded using a physically-based BRDF model~\cite{burley2012physically} based on the Trowbridge–Reitz (GGX) microfacet distribution~\cite{trowbridge75}. Results in Table~\ref{decomp-perf} indicate that rendering with NTC via stochastic filtering (see Section~\ref{sec:filtering_implemetation}) costs between 1.15~ms and 1.92~ms on a NVIDIA RTX 4090, while the cost decreases to 0.49~ms with traditional trilinear filtered BC7 textures.
The performance is similar for all materials in our evaluation set, and independent of the output channel count, ranging from three to twelve.
On the other hand, the varying number of features used across different compression profiles impacts the NTC performance. A higher number of features increases the sampling cost and the size of the network's first, input layer. We also implemented trilinear filtering for NTC by decompressing and filtering together eight texels and observed an $8\times$ slowdown.

Although NTC is more expensive than traditional hardware-accelerated texture filtering, our results demonstrate that our method achieves high performance and is practical for use in real-time rendering. 
Furthermore, when rendering a complex scene in a fully-featured renderer, we expect the cost of our method to be partially hidden by the execution of concurrent work (e.g., ray tracing) thanks to the GPU latency hiding capabilities.
The potential for latency hiding depends on various factors, such as hardware architecture, the presence of dedicated matrix-multiplication units that are otherwise under-utilized, cache sizes, and register usage.
We leave investigating this for future work.

\begin{table}[t]
\caption{Texture set quality as a function of material channel count.}
\vspace*{\tabvspace}
\resizebox{\columnwidth}{!}{%
\begin{tabular}{@{}r|c|c|c|c|c|c|c|c|c@{}}
\toprule
channels & 1 & 2 & 3 & 4 & 5 & 6 & 7 & 8 & 9 \\ \midrule
PSNR & 36.22 & 29.97 & 29.34 & 28.96 & 29.57 & 28.68 & 28.22 & 28.43 & 28.71 \\
BPPC & 2.19 & 1.1 & 0.73 & 0.55 & 0.44 & 0.36 & 0.31 & 0.27 & 0.244 \\ \bottomrule
\end{tabular}%
}
\label{channel_compression_effectiveness}
\end{table}

\section{Discussion}
In this section, we discuss various aspects of our compression. First, we verify the original motivation for our work by analyzing compression across channels and mipmap levels. Following this, we present various limitations of our approach, including failure cases. Finally, we discuss future work and potential applications.

\subsection{Compression Across Texture Channels and Mip Levels}
\emph{Texture Channels.}
Table~\ref{channel_compression_effectiveness}
shows results from our compression on a single texture set, with channel counts from 1 to 9.
We selected the \emph{Paving Stones} set as it is one of the most challenging texture set in our evaluation set.
We keep the compressed representation size fixed, expecting to observe a significant reduction in PSNR in case of uncorrelated properties, due to entropy limitations imposed by information theory.
Above two channels, we observe roughly constant quality, indicating the presence of significant cross-channel correlations, and suggesting that our model is able to effectively learn and exploit them.
The quality is not monotonic with respect to the channel count, because we report averaged PSNR across all channels.
Some channels are more challenging to compress than others, which leads to a PSNR increase when the introduced additional channel is easier to compress and more highly correlated with the previous ones. 
Overall, we achieve a similar low error across a large number of channels.

\emph{Mipmap Levels.}
Since our method shares a single decoder for all mipmap levels, we also analyzed
the impact of compressing only mipmap level 0 with the 0.2 BPPC profile,
and observed that the PSNR score was within 0.5 dB. This indicates a high level of feature reuse across mip levels.

\subsection{Limitations} \label{sec:limiations}
\begin{figure}[t]
\includegraphics[width=\linewidth]{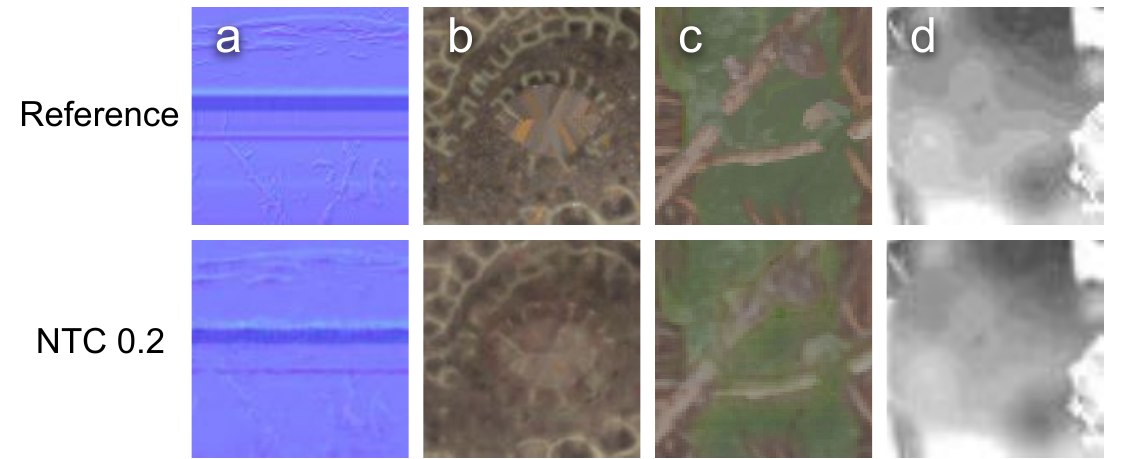}
\caption{Failure cases with our method. From left to right, a) removing fine details and noise, b) strong color shift, c) leaking of features between the texture channels, and d) removal of sharp staircase-like patterns.}
\centering
\label{fig:failure_cases}
\vspace{-1em}
\end{figure}

\emph{Failure Cases.}
Every lossy image or texture compression algorithm produces visual degradation %
at low bitrates. 
Typically, our method only results in mild blurring and color shifts. However, there are a few objectionable failure cases, which are presented in Figure~\ref{fig:failure_cases}.
We observed that two of the failure cases (c and d) result from potential material authoring errors,
namely misaligned texture channels and banding present in only a single channel, respectively, in
the reference uncompressed textures.
Our method relies heavily on channel correlation, and can be very sensitive to any alignment errors.
We include a more detailed discussion of these failure cases in our supplementary material (Appendix~\ref{supplement_compression_artifacts}).

\emph{Uniform Resolution.}
Our method relies on storing all textures within a single compressed material at the same resolution.
It is common practice for video game artists to store less visually important textures at smaller resolutions.
Our method can assign different levels of importance to textures by weighting its contribution to the loss, but otherwise requires all the single material input textures to be resampled to the same resolution before compression.

\emph{Distance-Dependent Benefits.}
Our method operates at different compression rate profiles, but one of the more interesting configurations is the one with lowest bitrate (NTC 0.2).
It demonstrates significantly more detail than BCx by enabling two higher resolution mipmap levels with the same storage. 
However, this increase in detail is not applicable at larger camera distances, when the additional mipmaps are no longer used.

\emph{Benefits Proportional to the Channel Count.}
Our method shows a high compression efficacy for materials with multiple channels.
However, for lower channel counts, e.g., just RGB textures, our storage cost is similar at iso-quality (Table~\ref{channel_compression_effectiveness}).
This means that our method would lose some of its advantage if it was to be applied to single textures or regular images.

\emph{Decompression of All Channels.}
Our method always decompresses \textit{all} material channels.
Only the last output layer can be simplified for extraction of fewer textures, and it does not significantly reduce cost.
This can be a limitation if different parts of  texture set are used in different rendering contexts, for example, a partial depth prepass that only requires opacity maps or tessellation that only accesses a displacement map. In such cases, it might be a better choice to compress these textures separately using traditional methods.

\emph{Filtering Cost.}
Unrolled filtering is computationally expensive, and stochastic filtering can introduce flickering by increasing the burden on spatiotemporal reconstruction~\cite{yang2020survey}.
Literature shows that it is possible to create filterable neural representations directly~\cite{kuznetsov2021neumip}, but we leave this for the future work.

\emph{Anisotropic Filtering.}
GPU texture samplers support \textit{anisotropic} filtering, which improves the appearance of objects in the distance.
However, a software implementation of anisotropic filtering with NTC would be prohibitively expensive for real-time rendering as it requires a large number of taps, each of which needs to be decoded.

\subsection{Future Work}
\emph{More Texture Types.}
Traditional GPU compression can support many types of textures, such as cube maps, 3D textures, and HDR textures.
We have not investigated the feasibility of application to those, and leave it for future work.

\emph{Appearance-Based Training.}
Recently emerging inverse and neural rendering techniques allow to use an appearance-based loss function. Using the rendering error to drive the texture compression instead of the BRDF property similarity could allow for even more efficient content adaptation.

\emph{Inlined Material Evaluation.}
Our method targets compression of arbitrary, generic material textures that can be used by an analytical or a neural renderer.
For the latter, there is no need to unpack materials to an intermediate representation.
It is possible to fuse together decompression and BRDF evaluation into a single MLP.
We leave this for future research.

\emph{Generative Textures and Materials.}
In our work, we target a faithful texture compression and preservation of existing detail.
It is possible to expand it further and use \textit{generative} approaches, where new plausible detail is generated upon zoom, despite not being present in the original texture.
Using some form of generative super-resolution, it should be possible to generate multiple finer additional texture mipmaps, without ever storing them on disk or in memory.

\emph{Further Optimizations.}
Further improvements to the compression ratio and inference speed could be achieved through lower precision intermediate computations.

\section{Conclusion}
We have introduced a novel texture compression algorithm, targeting the increasing memory and fidelity requirements of modern computer graphics applications, 
and new, richer physically-based shading models that require many properties, commonly stored in textures.
For high-performance texture accesses, it is of utmost importance to be able to spatially
access the textures anywhere at a small cost, which is often referred to as the
random access property. %
We have shown that very high compression rates can be achieved even without sacrificing local and random access.

By compressing many channels and mipmap levels together, the quality of our algorithm's low bitrate results surpasses that of state-of-the-art industry standards, such as JPEG XL and AVIF that are substantially more complex methods, without requiring entropy coding.

By utilizing matrix multiplication intrinsics available in the off-the-shelf GPUs, 
we have shown that decompression of our textures 
introduces only a modest timing overhead as compared to simple BCx algorithms (which executes in custom hardware), possibly making our method practical in disk- and memory-constrained graphics applications.

We hope our work will inspire the creation of highly compressed neural representations for use in other areas of real-time rendering, as a means of achieving cinematic quality.

\begin{figure*}[ht]
	\setlength{\tabcolsep}{1.0pt}%
	\renewcommand{\arraystretch}{0.7}
	\newcommand{\imgW}{0.13\textwidth}
	\newcommand{\linelen}{36.42mm}
        \newcommand{\raiseboxheight}{3.25mm}
	\centering
	{\footnotesize
		\begin{tabular}{cccccccc}
			& & & & \multicolumn{4}{c}{\line(1,0){\linelen}{\normalsize\  0.2 BPPC\ }\line(1,0){\linelen}} \\
			& & Original & Original & BC high & AVIF & JPEG XL & NTC \\
			\multirow{2}{*}[5mm]{\rotatebox{90}{{\normalsize diffuse map}}}
			& \raisebox{\raiseboxheight}{\rotatebox{90}{mipmap level 0}} &
			\includegraphics[width=\imgW]{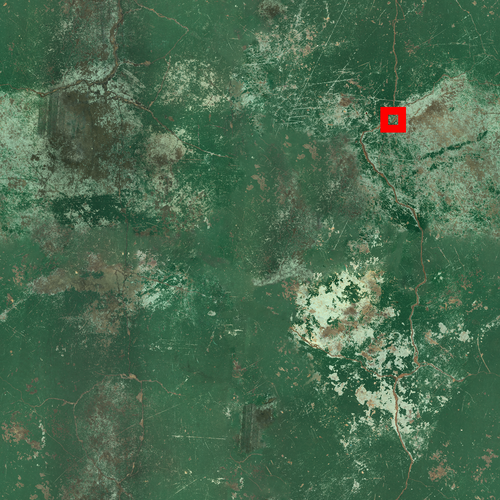} &
			\includegraphics[width=\imgW]{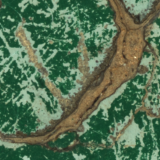} &
			\includegraphics[width=\imgW]{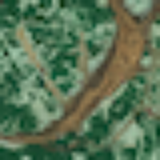} &
			\includegraphics[width=\imgW]{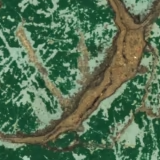} &
			\includegraphics[width=\imgW]{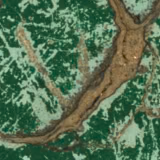} &
			\includegraphics[width=\imgW]{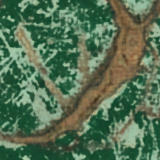} \\
			& \raisebox{\raiseboxheight}{\rotatebox{90}{mipmap level 3}} &
			\includegraphics[width=\imgW]{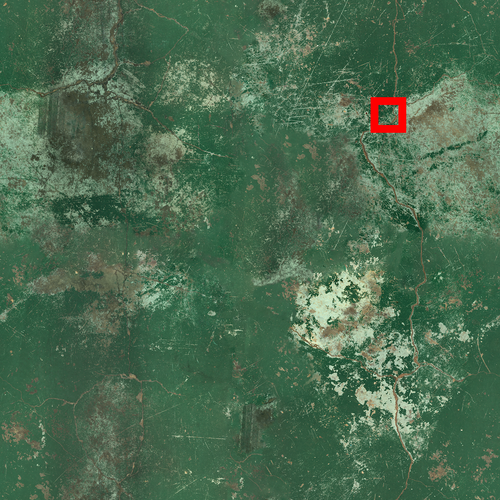} &
			\includegraphics[width=\imgW]{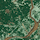} &
			\includegraphics[width=\imgW]{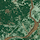} &
			\includegraphics[width=\imgW]{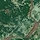} &
			\includegraphics[width=\imgW]{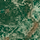} &
			\includegraphics[width=\imgW]{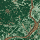} 
			\\
			\hline \vspace*{-1mm}
			\\
			\multirow{2}{*}[6mm]{\rotatebox{90}{{\normalsize normal map}}}
			& \raisebox{\raiseboxheight}{\rotatebox{90}{mipmap level 0}} &
			\includegraphics[width=\imgW]{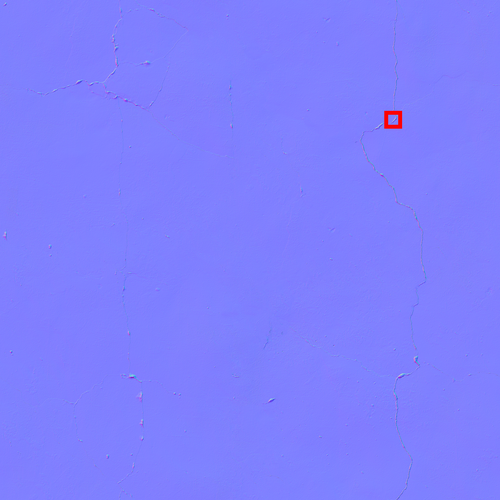} &
			\includegraphics[width=\imgW]{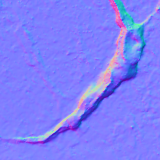} &
			\includegraphics[width=\imgW]{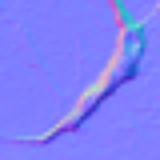} &
			\includegraphics[width=\imgW]{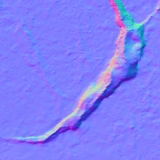} &
			\includegraphics[width=\imgW]{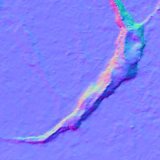} &
			\includegraphics[width=\imgW]{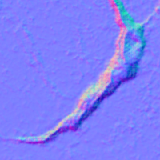} \\
			& \raisebox{\raiseboxheight}{\rotatebox{90}{mipmap level 3}} &
			\includegraphics[width=\imgW]{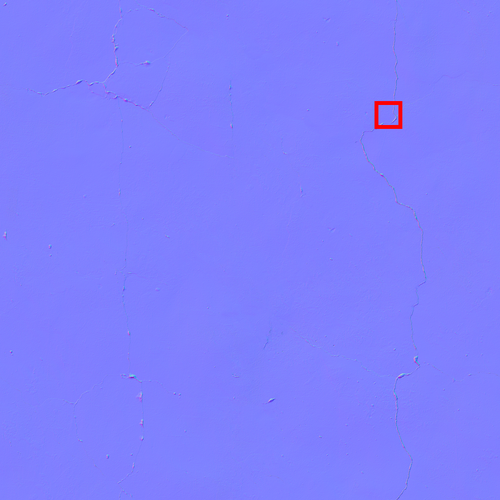} &
			\includegraphics[width=\imgW]{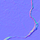} &
			\includegraphics[width=\imgW]{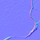} &
			\includegraphics[width=\imgW]{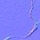} &
			\includegraphics[width=\imgW]{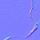} &
			\includegraphics[width=\imgW]{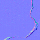}
			\\
			\hline \vspace*{-1mm}
			\\
			\multirow{2}{*}[10mm]{\rotatebox{90}{{\normalsize displacement map}}}
			& \raisebox{\raiseboxheight}{\rotatebox{90}{mipmap level 0}} &
			\includegraphics[width=\imgW]{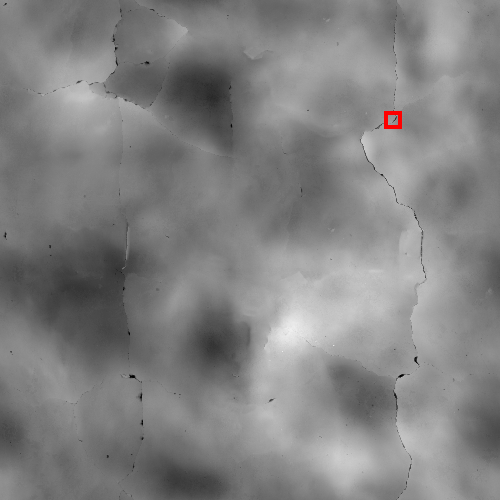} &
			\includegraphics[width=\imgW]{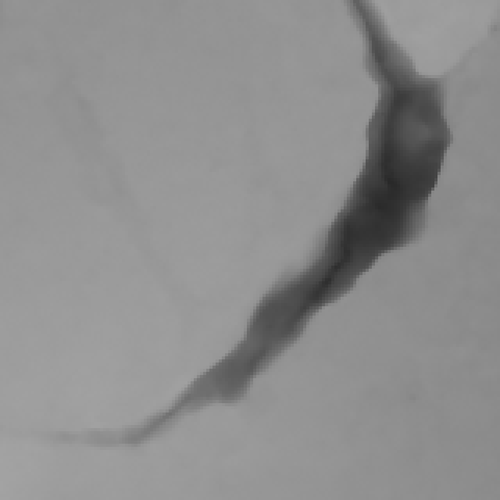} &
			\includegraphics[width=\imgW]{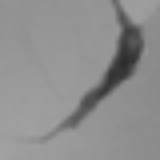} &
			\includegraphics[width=\imgW]{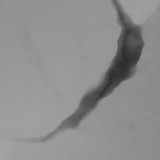} &
			\includegraphics[width=\imgW]{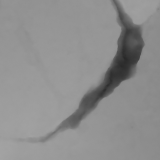} &
			\includegraphics[width=\imgW]{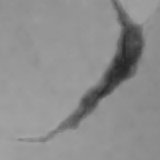} \\
			& \raisebox{\raiseboxheight}{\rotatebox{90}{mipmap level 3}} &
			\includegraphics[width=\imgW]{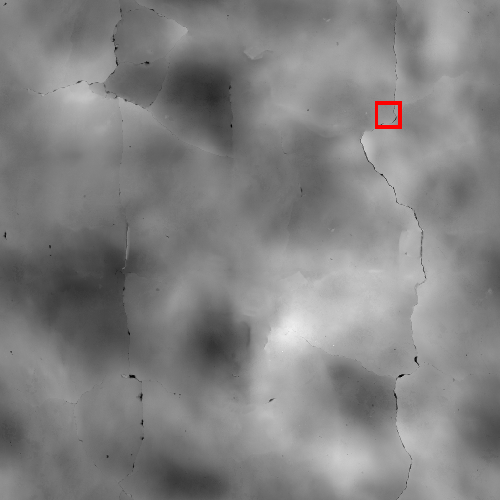} &
			\includegraphics[width=\imgW]{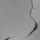} &
			\includegraphics[width=\imgW]{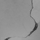} &
			\includegraphics[width=\imgW]{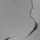} &
			\includegraphics[width=\imgW]{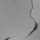} &
			\includegraphics[width=\imgW]{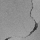}
			\\
			\hline \vspace*{-1mm}
			\\
			\multirow{2}{*}[2mm]{\rotatebox{90}{{\normalsize ARM}}}
			& \raisebox{\raiseboxheight}{\rotatebox{90}{mipmap level 0}} &
			\includegraphics[width=\imgW]{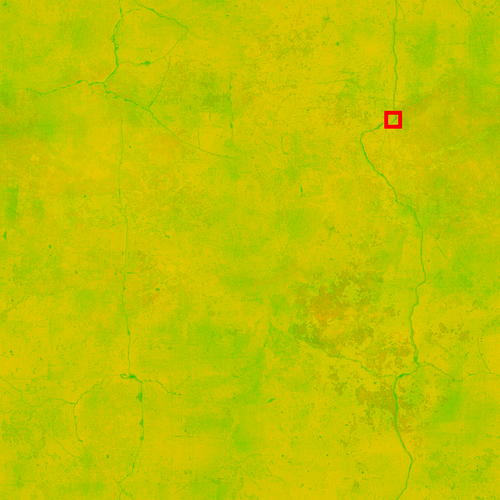} &
			\includegraphics[width=\imgW]{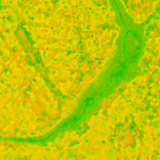} &
			\includegraphics[width=\imgW]{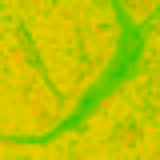} &
			\includegraphics[width=\imgW]{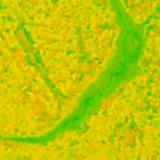} &
			\includegraphics[width=\imgW]{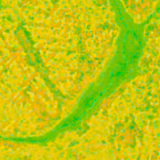} &
			\includegraphics[width=\imgW]{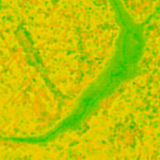} \\
			& \raisebox{\raiseboxheight}{\rotatebox{90}{mipmap level 3}} &
			\includegraphics[width=\imgW]{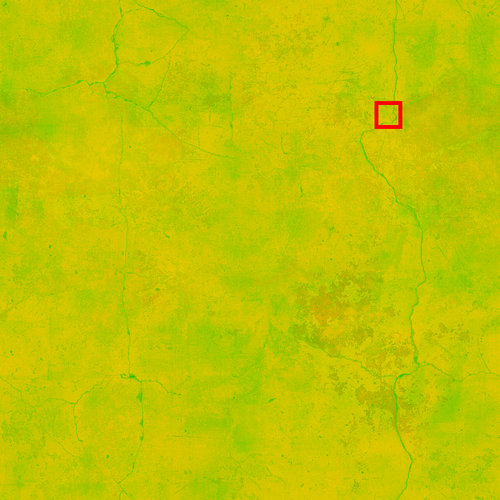} &
			\includegraphics[width=\imgW]{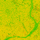} &
			\includegraphics[width=\imgW]{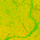 } &
			\includegraphics[width=\imgW]{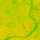} &
			\includegraphics[width=\imgW]{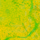} &
			\includegraphics[width=\imgW]{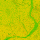}
		\end{tabular}
	}
	\caption{Comparison of different methods at 0.2~BPPC, where we selected to show
		the texture set for which NTC's PSNR was closest to its average PSNR over all texture sets in our evaluation dataset. Recall that neither AVIF nor JPEG XL provide random access to the texture data. Furthermore,
        to achieve an approximately iso-storage comparison,
        mipmap level 0 images for BC high were created by
         bilinear upsampling of its mipmap level 2.
            The textures were retrived from \url{https://polyhaven.com/}.
            Additional examples are available in our supplementary material.
		}
	\label{fig_big_comparison}
\end{figure*}

\begin{acks}
We are grateful for the help of our colleagues on this project: John Burgess, for many valuable suggestions and ideas; Yong He, Patrick Neill, Justin Holewinski, and Petrik Clarberg for their help with Slang programming language, driver support for fast matrix multiplication, and performance evaluation; Toni Bratincevic and Andrea Weidlich for the Inkwell asset and its material adapted to the Falcor framework.
Finally, we would like to thank Lennart Demes, author of the ambientCG website, for providing a public domain material database that we used to evaluate the efficacy of our algorithm.
This work was partially supported by the Wallenberg AI, Autonomous Systems and Software
Program (WASP) funded by the Knut and Alice Wallenberg Foundation. %
\end{acks}

\bibliographystyle{ACM-Reference-Format}
\bibliography{bibliography}

\appendix

\vfill\null
\section{Handling Divergence}
\label{handling_divergence}

Using matrix acceleration for the neural network requires all SIMD lanes to be active and network weights to be uniform across the SIMD lanes. However, in some scenarios like ray tracing, rays from the same SIMD group may hit different materials or miss geometry altogether. When querying ray-scene intersections from a compute shader, users can control the execution mask, ensuring all SIMD lanes are active during network evaluation. Conversely, in hit or miss shaders, users lack control over the shader execution mask. In these cases, the users can query the execution mask and enable tensor acceleration when all lanes are active, otherwise a fallback path without tensor acceleration is necessary.

The following example code shows how divergence can be handled inside a hit shader by enabling matrix acceleration when all lanes are active, and by iterating over unique sets of network parameter offsets, which are broadcast across all SIMD lanes to make them uniform. SIMD occupancy in a complex scene with a large number of materials can potentially be improved with techniques like SER~\shortcite{Ser} and TSU~\shortcite{Tsu}. We leave this evaluation for future work.

\lstset{emph={uint, bool},emphstyle={\color{blue}}}

\begin{lstlisting}[language=C]
Outputs runNetwork(Inputs x, uint paramOffsets) {
  // Check if all lanes are active.
  if (WaveActiveCountBits(true) == WaveGetLaneCount()) { 
    uint mask = -1;
    uint lane = 0; 
    // Iterate over unique network parameters in the SIMD group.
    for (; mask ;) {
      // Broadcast the parameter offset across SIMD lanes.
      uint offset = WaveReadLaneAt(paramOffsets, lane); 
      bool matchingLanes = offset == paramOffsets;

      // Evaluate the MLP with matrix acceleration.
      Outputs y = MLP(x, offset);

      // Store the outputs for matching lanes.
      storeOutputs(y, matchingLanes);
      
      // Clear the evaluated lanes.
      mask -= WaveActiveBallot(matchingLanes).x;
      lane = firstbitlow(mask);            
    }
  } else {
    // Fallback without matrix acceleration.
  }
}

\end{lstlisting}

\clearpage

\begin{figure*}[ht]
	\centering
	\newcommand{\imgH}{0.217\textwidth}
	\setlength{\tabcolsep}{1.0pt}%
        {\scriptsize
    	\begin{tabular}{ccccc}
            & \textbf{BC high}. PNSR (\textuparrow): 29.1~dB, \FLIP (\textdownarrow): 0.14   
            & \textbf{NTC}. PSNR (\textuparrow): \textbf{36.0}~dB, \FLIP (\textdownarrow): \textbf{0.08}
            & \textbf{reference}: not compressed
            \\
            & $1024\times 1024$ at 4.0~MB.
            & $4096\times 4096$ at \textbf{3.8}~MB.
            & $4096\times 4096$ at 171~MB.
            \\
    		\includegraphics[height=\imgH]{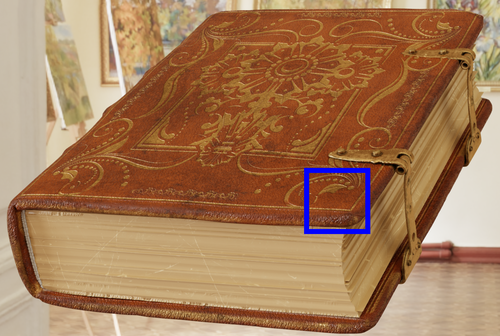} &
    		\includegraphics[height=\imgH]{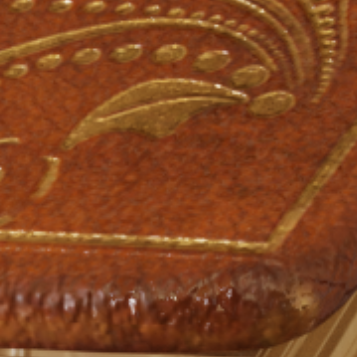} &
    		\includegraphics[height=\imgH]{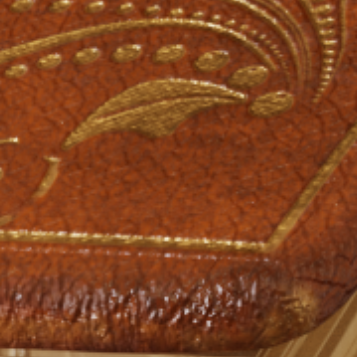} &
                \includegraphics[height=\imgH]{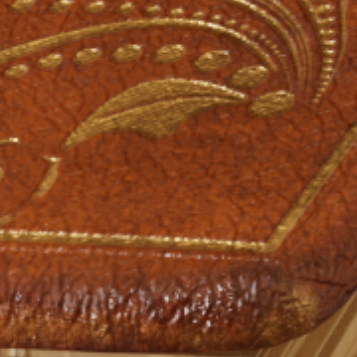}
    		\vspace*{-39.6mm}		
                \\
    		\vspace*{26.5mm}		
        		& 
        		\hspace*{29.5mm}
        		\includegraphics[height=0.05\textwidth]{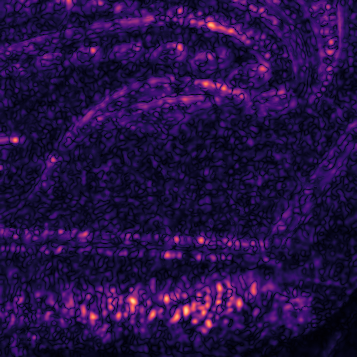} &
        		\hspace*{29.4mm}
        		\includegraphics[height=0.05\textwidth]{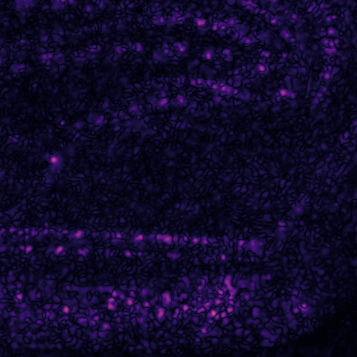}
                &
                &
                \hspace*{-1.5mm}
                \raisebox{0mm}{\rotatebox{90}{\includegraphics[width=0.05\textwidth, height=1mm]{images/magma.png}}}
                        \\
    	\end{tabular}
            }
            \caption{A rendered image of a closed book. The cutouts demonstrate quality using, from
        left to right, GPU-based texture formats (BC high) at $1024\times 1024$ resolution,
        our neural texture compression (NTC) , and high-quality reference textures. 
        Note that NTC provides two additional mipmap levels over BC high, despite it using
        slightly less memory.
        The metrics, PSNR and \FLIP, were computed for the cutouts and are shown above the respective image.
        The \FLIP error images, whose brightness is proportional to error, are shown in the upper right corners.
  }
  \Description{TODO}
  \label{fig:closed_book}
\end{figure*}

\begin{figure*}[ht]
	\centering
	\newcommand{\imgH}{0.217\textwidth}
	\setlength{\tabcolsep}{1.0pt}%
        {\scriptsize
    	\begin{tabular}{ccccc}
                & \textbf{BC high}. PNSR (\textuparrow): 22.5~dB, \FLIP (\textdownarrow): 0.27   
			& \textbf{NTC}. PSNR (\textuparrow): \textbf{35.8}~dB, \FLIP (\textdownarrow): \textbf{0.11}
			& \textbf{reference}: not compressed
                \\
                & $1024\times 1024$ at 4.0~MB.
                & $4096\times 4096$ at \textbf{3.8}~MB.
                & $4096\times 4096$ at 171~MB.
                \\
    		\includegraphics[height=\imgH]{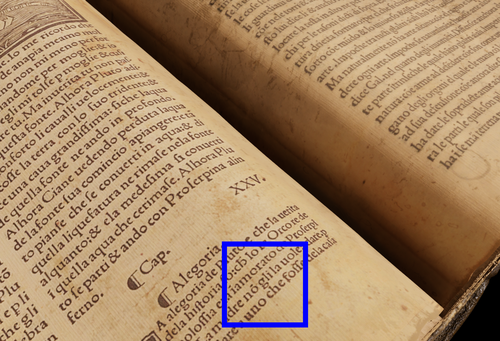} &
    		\includegraphics[height=\imgH]{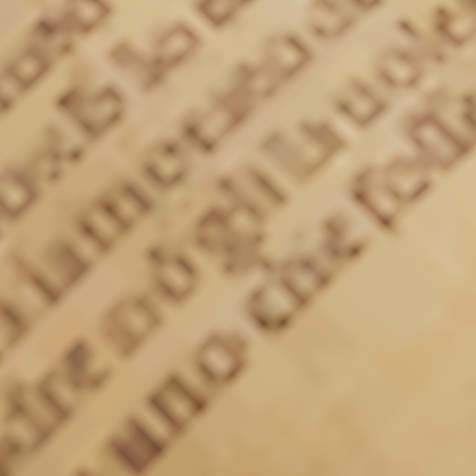} &
    		\includegraphics[height=\imgH]{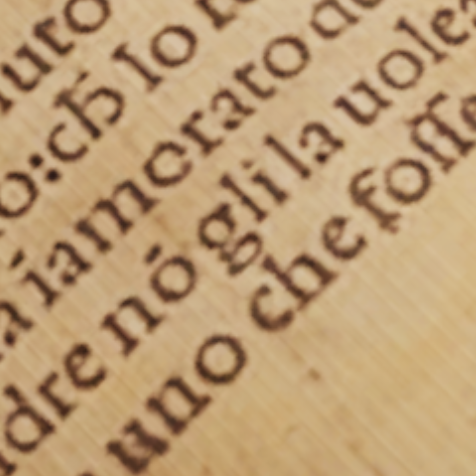} &
                \includegraphics[height=\imgH]{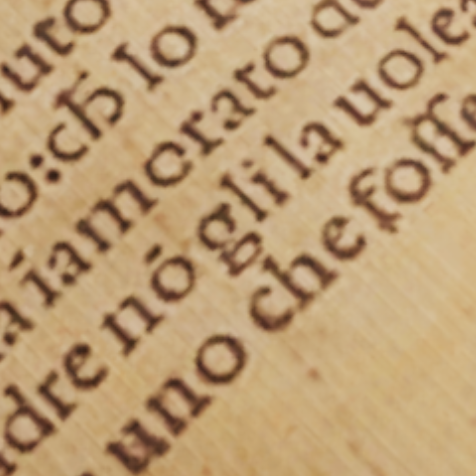}
    		\vspace*{-39.6mm}		
                \\
    		\vspace*{26.5mm}		
        		& 
        		\hspace*{29.5mm}
        		\includegraphics[height=0.05\textwidth]{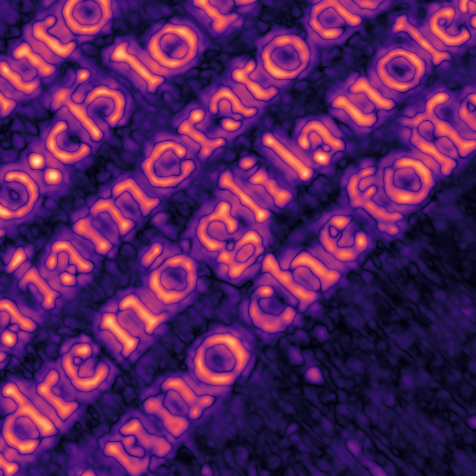} &
        		\hspace*{29.4mm}
        		\includegraphics[height=0.05\textwidth]{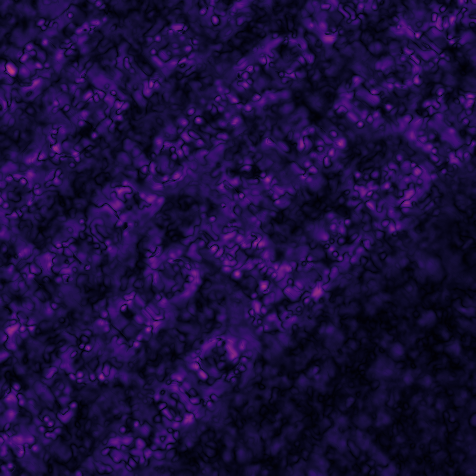}
                &
                &
                \hspace*{-1.5mm}
                \raisebox{0mm}{\rotatebox{90}{\includegraphics[width=0.05\textwidth, height=1mm]{images/magma.png}}}
    	\end{tabular}
            }
  \caption{A rendered image of an open book. See Figure~\ref{fig:closed_book}'s caption for details.
  }
  \Description{TODO}
  \label{fig:open_book}
\end{figure*}

\section*{\textbf{Supplementary to \papertitle}}

\section{Rendered Quality}
\label{sec:rendered_images_examples}
In addition to the first figure in the main paper,
Figures~\ref{fig:closed_book} and \ref{fig:open_book} show examples of rendered images
using high-quality non\hyp{}compressed textures, %
textures compressed with our method (using the lowest BPPC profile, NTC~0.2),
and textures compressed with BC high. For the latter, the two highest-resolution mip levels were created
through bilinear upscaling of the third mip level in order to obtain an iso-storage comparison to our method.
The rendered images, supported by the error images and values,
show that NTC achieves higher quality rendered images than the iso-storage version of BC high.
This is especially noticeable in Figure~\ref{fig:open_book},
which also gives an indication that NTC does well on textures with text.

\section{Additional Quantitative Results} \label{app:quantitative}
The following two subsections demonstrate how the quality
of NTC's compressed images changes over the mipmap chain (Section~\ref{sec:mipmap_quality})
and how it differs for various texture types (Section~\ref{sec:texture_type_quality}).

\subsection{Mipmap Quality}
\label{sec:mipmap_quality}

The performance of a compression technique can vary based on the frequency spectrum of the image
and therefore perform differently across mip levels.
In Figure~\ref{fig:mip_quality_plot}, we compare the per-mip-level PSNR scores of NTC~0.2 to the BC high algorithm.
Since there is a $16\times$ difference in storage cost between the algorithms, an iso-storage comparison was conducted, resulting in the omission of the first two mip levels for BC compression.
PSNR values with our method are either comparable to, or higher than BC depending on the mip level,
except for mip levels two and three where NTC shows slightly worse PSNR scores.

\subsection{Different Texture Type Compression Quality}
\label{sec:texture_type_quality}

\begin{figure}[t]
\scalebox{0.5}{\input{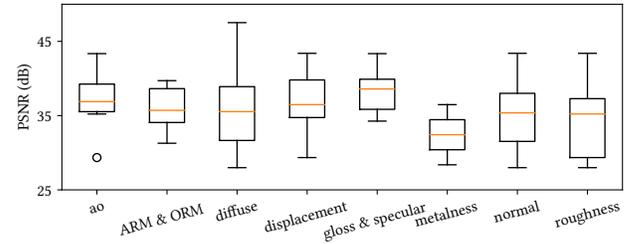}}
\caption{Compression quality of different material properties for NTC~0.2. The orange lines show the median PSNR value for the respective texture types.}
\centering
\label{fig:texture_type_quality}
\end{figure}

Figure~\ref{fig:texture_type_quality} presents the PSNR scores of NTC 0.20, computed on different types of textures in the material texture set, such as diffuse, normals, etc. Given the similarities in data content between certain texture types, such as ARM and ORM textures, as well as gloss and specular textures, the results for these pairs were concatenated. Additionally, texture types that occurred only once in our data set (see Section~\ref{sec_texture_set_selection})  were excluded. 
The results indicate that our proposed method is able to compress different texture types at similar levels of quality.

\section{Compression artifacts}
\label{supplement_compression_artifacts}
\begin{figure}[b]
\includegraphics[width=\linewidth]{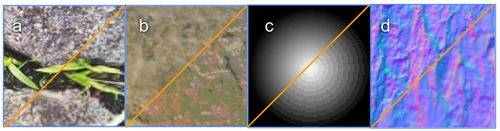}
\caption{Examples of typical compression artifacts. \textbf{a)} visible blocks \textbf{b)} posterization and discoloration \textbf{c)} gradient banding \textbf{d)} detail loss and bluriness. }
\label{fig:fig_artifacts}
\end{figure}

Every texture compression algorithm degrades quality differently.
Particularly visible quality differences are called \textit{compression artifacts} and we present a few typical examples in Figure~\ref{fig:fig_artifacts}. 
Block-based compression methods commonly exhibit visible block artifacts (inset a). 
Methods that rely on heavy quantization tend to exhibit banding artifacts, as illustrated in insets (b) and (c), and are often characterized by a visible discoloration towards green or purple hues, resulting from higher chroma quantization (inset b).
These artifacts are highly perceptible to the human eye, and modern image compression techniques such as AVIF and JPEG XL, have prioritized their removal, by producing blurry images instead (inset d).

\begin{figure}[t]
\centering
\begin{tabular}{cc}
     \hspace*{-1.25mm}\includegraphics[width=0.95\linewidth]{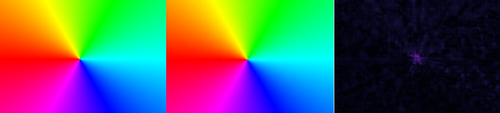} &
     \hspace*{-3mm}\raisebox{0mm}{\rotatebox{90}{\includegraphics[width=0.2175\linewidth, height=0.025\linewidth]{images/magma.png}}}
\end{tabular}
\caption{A colorful and (presumably) difficult gradient texture compressed with NTC 0.2 does not show visible banding, color posterization, or discoloration. \textbf{Left:} Reference. \textbf{Middle:} Compressed. \textbf{Right:} \FLIP error image and corresponding color map.
}
\label{fig:fig_gradient_no_artifacts}
\end{figure}

Since our feature vectors are quantized down to \textit{two} or \textit{four} bits per feature, we specifically check for the presence of banding artifacts by compressing a synthetic gradient texture and do not observe noticeable banding artifacts as shown in Figure~\ref{fig:fig_gradient_no_artifacts}. We attribute this to the combination of smooth, bilinear interpolation and the higher frequency learned interpolation (see Sections~\ref{pyramid} and~\ref{sec_sampling_concatenation}).

Figure~\ref{fig_big_comparison} demonstrates the absence of visually objectionable artifacts with our compression method on an average case.
In contrast, we observe the presence of block artifacts on the ARM texture with BC, especially for mip level 0.
AVIF and JPEG XL produce sharper results than NTC for the diffuse texture at mip level 0 but are significantly blurrier or show some discoloration (JPEG XL) at mip level 3. 
This is likely because with these methods, different mip levels are compressed separately, and their spectral content does not necessarily reduce proportionally to the resolution. 
These observations make a strong case for jointly compressing mip levels as we do with NTC.
Potentially, we could compress AVIF and JPEG XL by allocating different rates for each mip as well as each texture, while maintaining the overall storage constant. However, determining suitable rates can be challenging, particularly as it can be material specific.
We do not include comparisons with heterogeneous rates for this reason, and also because our proposed method does not aim to compete directly with these image compression techniques.

On average, our method produces results that are a bit blurrier and sometimes less saturated than the uncompressed reference, but significantly better than BCx compression.
There are, however, some more objectionable failure cases presented in Figure~\ref{fig:failure_cases}
in the main paper.
In example a) we observe strong distortion of the normal map of the \emph{Ticket Machine} texture.
The compressed texture is very flat, mostly sparse, and our optimization procedure fails to reconstruct subtle details properly.
In example b) (albedo of the \emph{dragon atlas} material), we observe discoloration of the texture.
Since our approach is specialized for each material, it can adapt to high frequency content, such as detailed normal maps, or large color variations.
However, the most challenging materials have both kinds of features.
In such cases, we cannot reconstruct both features equally well given the low BPPC rates.
Typically, we observe that details and normal map features are favored by our optimization because of their higher variance, at the cost of other material textures. 
It is possible to balance the quality of material textures by adjusting the loss function (Section~\ref{optimization_and_loss}). 
In scenarios where the material textures have a fixed set of semantics, we can apply more robust texture specific optimization, such as a loss in chrominance space, or optimization based on appearance using differentiable renderering.
The last two failure cases (c and d) in Figure~\ref{fig:failure_cases} correspond to unusual data in the source materials.
In example c) (\emph{Pine Forest Ground} texture), the normal maps seem to be misaligned with the albedo maps, producing leakage of details between channels.
Example d) (metalness map of the \emph{Metal Plates} texture) shows strong banding in the source (reference) texture, present only in a single material channel, and our method blurs it and correlates with the other material channels.

\begin{figure*}[h!]
	\setlength{\tabcolsep}{1.0pt}%
	\renewcommand{\arraystretch}{0.7}
	\newcommand{\imgW}{0.1\textwidth}
	\newcommand{\linelen}{30.35mm}
        \newcommand{\raiseboxheight}{6.5mm}
	\centering
	{\footnotesize
		\begin{tabular}{cccccccc}
			& & & & \multicolumn{4}{c}{\line(1,0){\linelen}{\normalsize\  0.5 BPPC\ }\line(1,0){\linelen}} \\
			& & Original & Original & ASTC $12\times 12$ & AVIF & JPEG XL & NTC \\
			\multirow{2}{*}[5mm]{\rotatebox{90}{{\normalsize diffuse map}}}
			& \raisebox{\raiseboxheight}{\rotatebox{90}{mip 0}} &
			\includegraphics[width=\imgW]{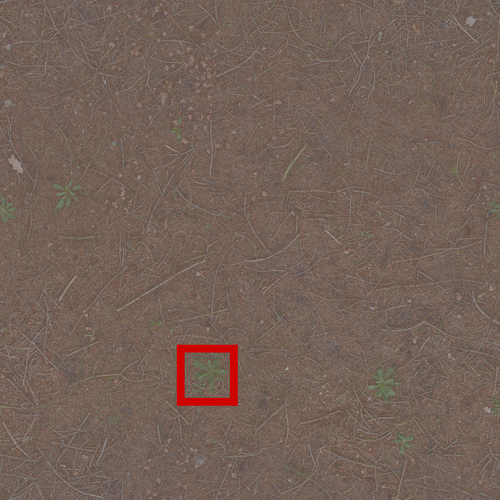} &
			\includegraphics[width=\imgW]{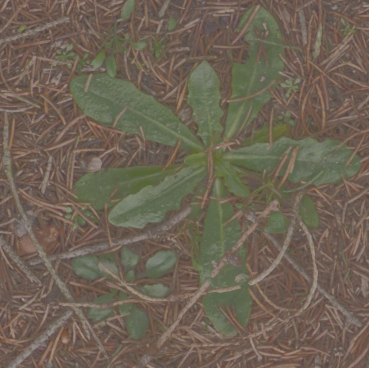} &
			\includegraphics[width=\imgW]{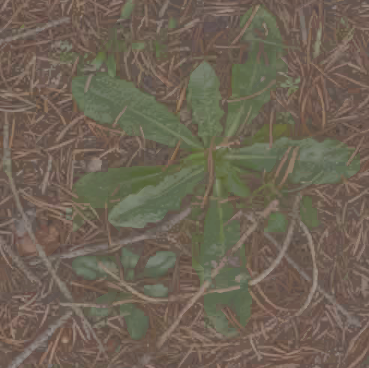} &
			\includegraphics[width=\imgW]{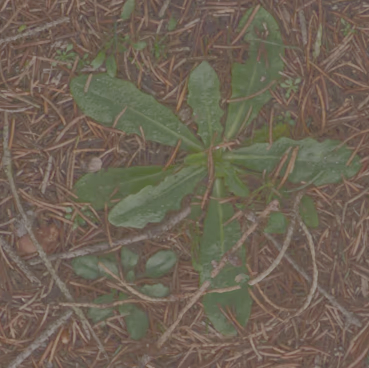} &
			\includegraphics[width=\imgW]{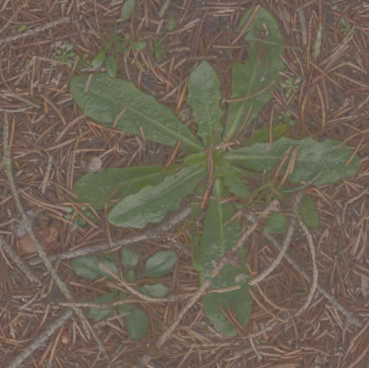} &
			\includegraphics[width=\imgW]{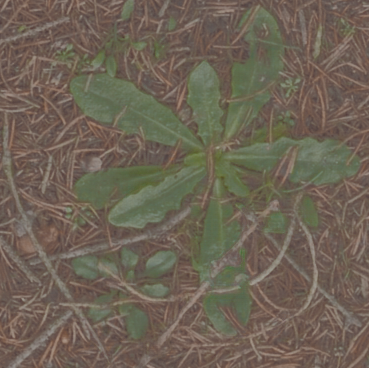} \\
			& \raisebox{\raiseboxheight}{\rotatebox{90}{mip 3}} &
			\includegraphics[width=\imgW]{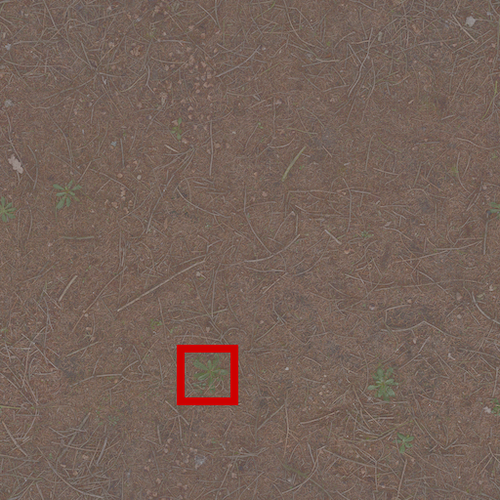} &
			\includegraphics[width=\imgW]{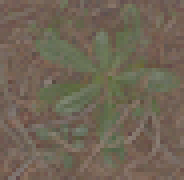} &
			\includegraphics[width=\imgW]{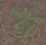} &
			\includegraphics[width=\imgW]{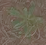} &
			\includegraphics[width=\imgW]{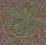} &
			\includegraphics[width=\imgW]{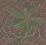} 
			\\
			\hline \vspace*{-1mm}
			\\
			\multirow{2}{*}[6mm]{\rotatebox{90}{{\normalsize normal map}}}
			& \raisebox{\raiseboxheight}{\rotatebox{90}{mip 0}} &
			\includegraphics[width=\imgW]{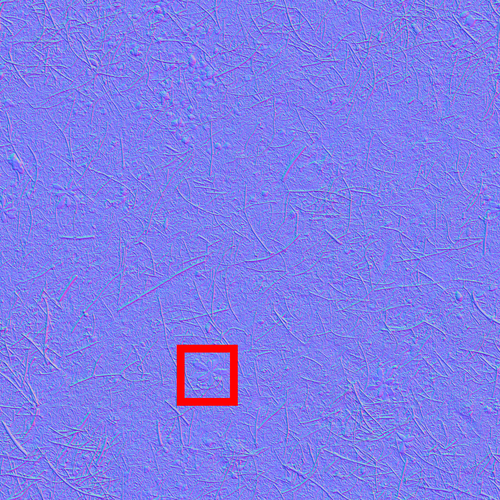} &
			\includegraphics[width=\imgW]{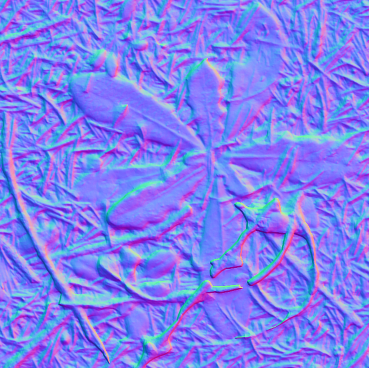} &
			\includegraphics[width=\imgW]{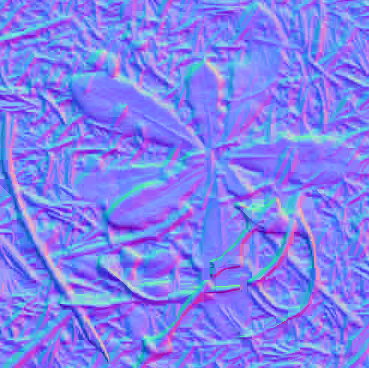} &
			\includegraphics[width=\imgW]{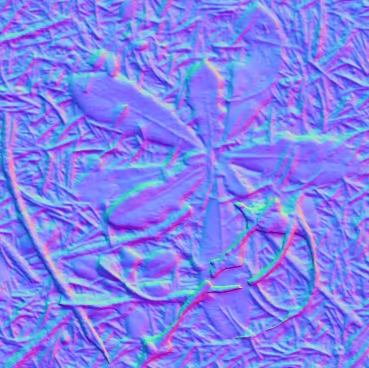} &
			\includegraphics[width=\imgW]{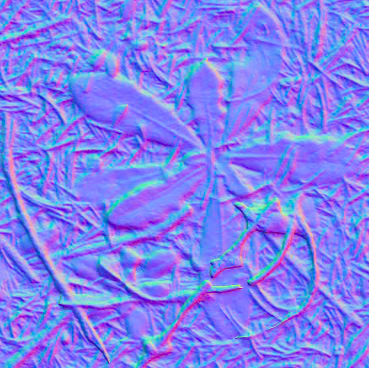} &
			\includegraphics[width=\imgW]{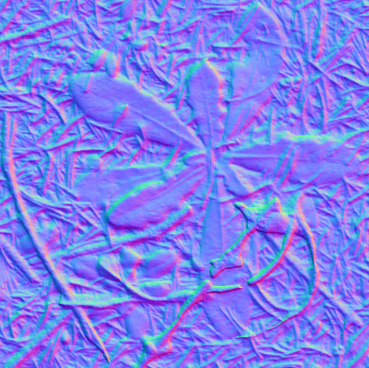} \\
			& \raisebox{\raiseboxheight}{\rotatebox{90}{mip 3}} &
			\includegraphics[width=\imgW]{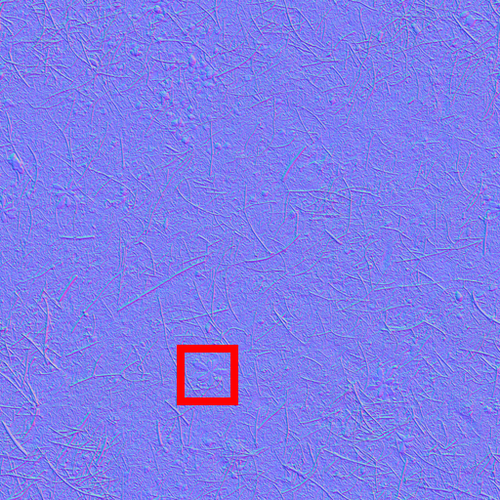} &
			\includegraphics[width=\imgW]{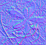} &
			\includegraphics[width=\imgW]{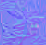} &
			\includegraphics[width=\imgW]{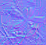} &
			\includegraphics[width=\imgW]{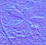} &
			\includegraphics[width=\imgW]{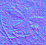}
			\\
			\hline \vspace*{-1mm}
			\\
			\multirow{2}{*}[10mm]{\rotatebox{90}{{\normalsize displacement map}}}
			& \raisebox{4.5mm}{\rotatebox{90}{mip 0}} &
			\includegraphics[width=\imgW]{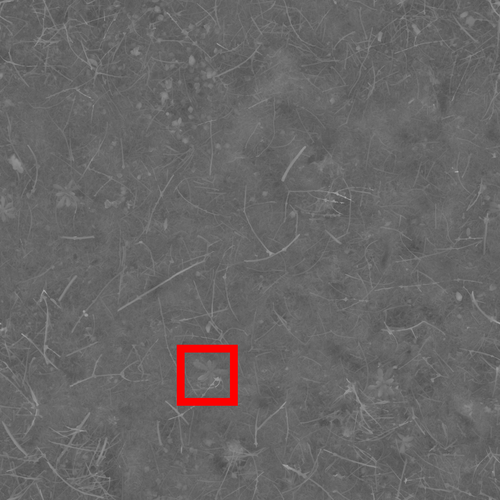} &
			\includegraphics[width=\imgW]{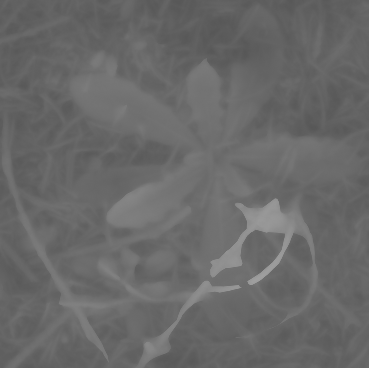} &
			\includegraphics[width=\imgW]{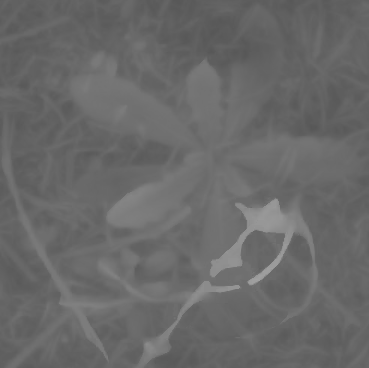} &
			\includegraphics[width=\imgW]{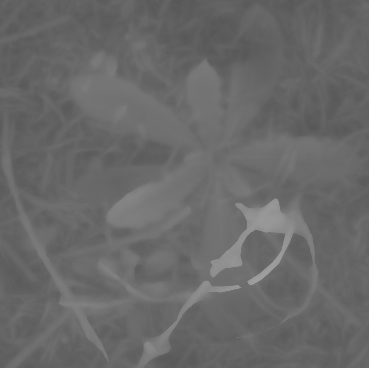} &
			\includegraphics[width=\imgW]{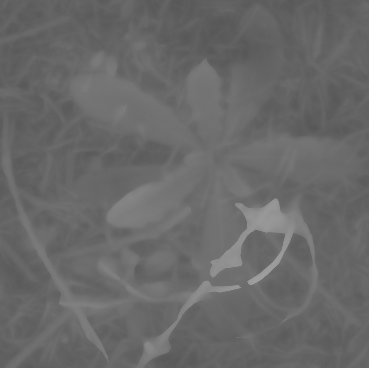} &
			\includegraphics[width=\imgW]{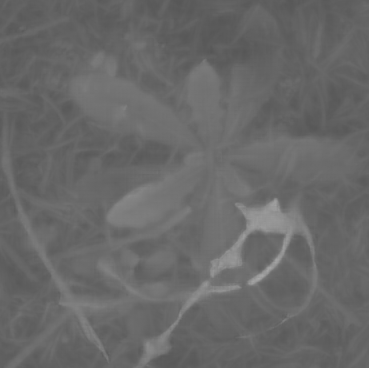} \\
			& \raisebox{\raiseboxheight}{\rotatebox{90}{mip 3}} &
			\includegraphics[width=\imgW]{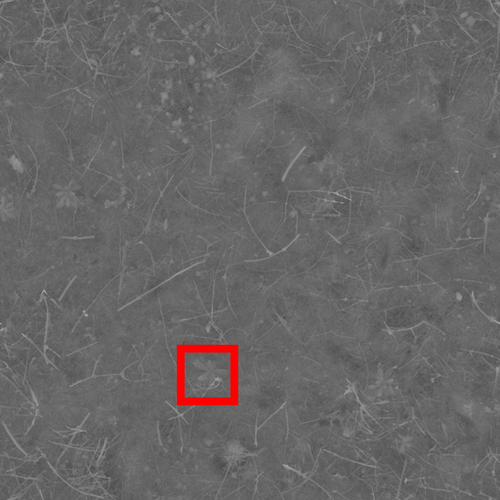} &
			\includegraphics[width=\imgW]{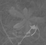} &
			\includegraphics[width=\imgW]{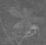} &
			\includegraphics[width=\imgW]{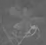} &
			\includegraphics[width=\imgW]{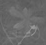} &
			\includegraphics[width=\imgW]{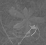}
			\\
			\hline \vspace*{-1mm}
			\\
			\multirow{2}{*}[8mm]{\rotatebox{90}{{\normalsize specular map}}}
			& \raisebox{6mm}{\rotatebox{90}{mip 0}} &
			\includegraphics[width=\imgW]{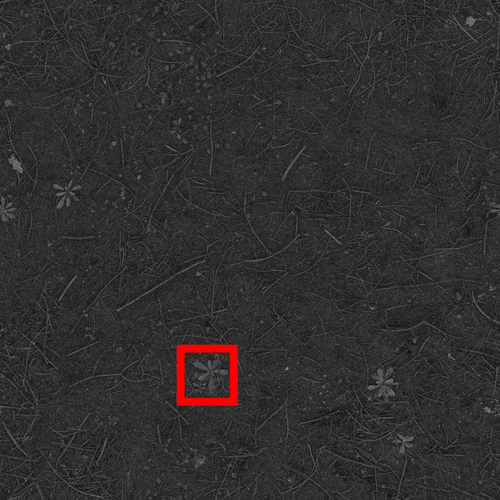} &
			\includegraphics[width=\imgW]{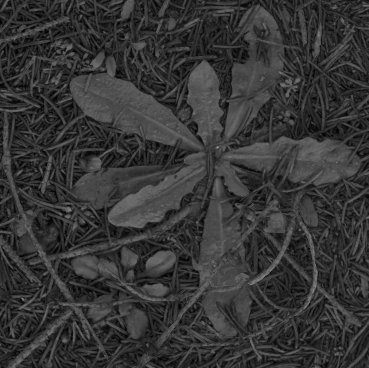} &
			\includegraphics[width=\imgW]{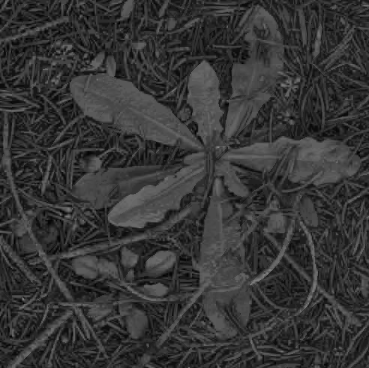} &
			\includegraphics[width=\imgW]{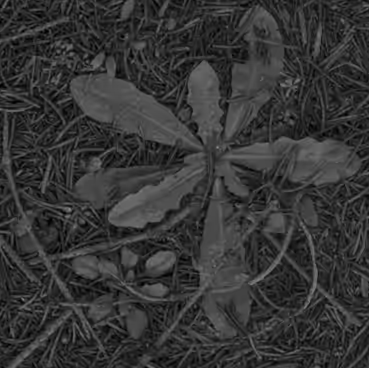} &
			\includegraphics[width=\imgW]{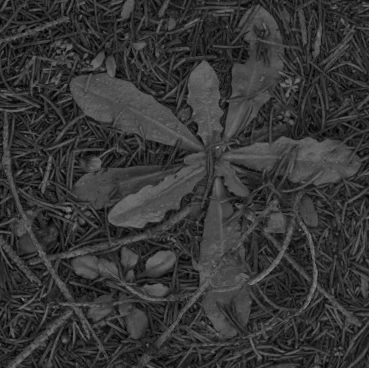} &
			\includegraphics[width=\imgW]{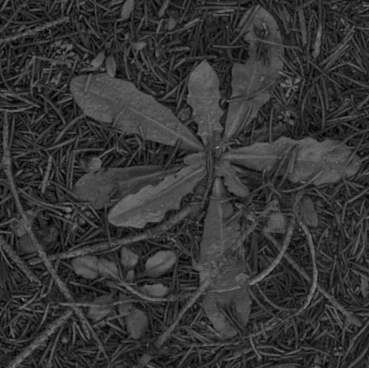} \\
			& \raisebox{\raiseboxheight}{\rotatebox{90}{mip 3}} &
			\includegraphics[width=\imgW]{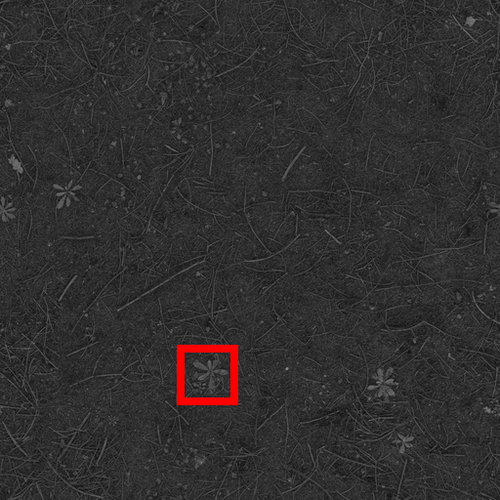} &
			\includegraphics[width=\imgW]{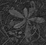} &
			\includegraphics[width=\imgW]{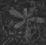 } &
			\includegraphics[width=\imgW]{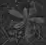} &
			\includegraphics[width=\imgW]{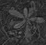} &
			\includegraphics[width=\imgW]{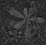}
                \\
                \hline \vspace*{-1mm}
			\\
			\multirow{2}{*}[7mm]{\rotatebox{90}{{\normalsize roughness map}}}
			& \raisebox{\raiseboxheight}{\rotatebox{90}{mip 0}} &
			\includegraphics[width=\imgW]{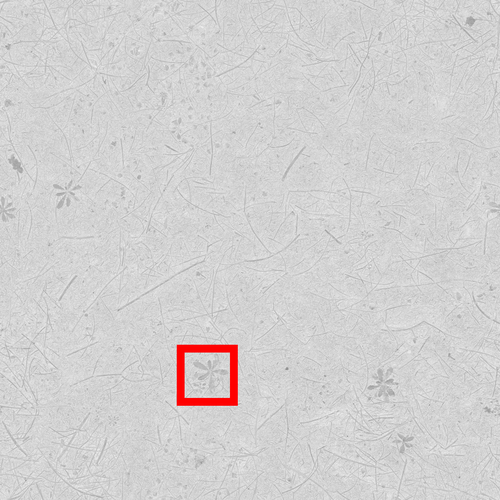} &
			\includegraphics[width=\imgW]{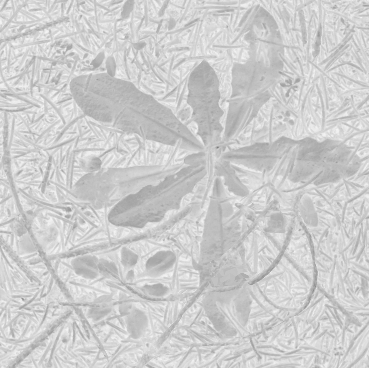} &
			\includegraphics[width=\imgW]{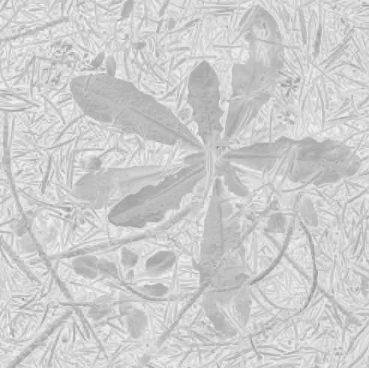} &
			\includegraphics[width=\imgW]{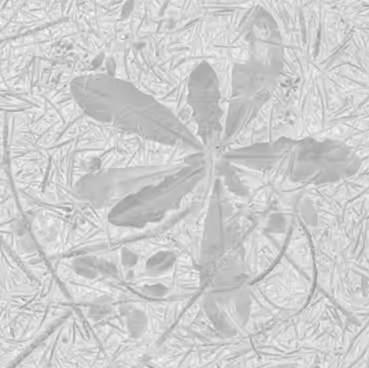} &
			\includegraphics[width=\imgW]{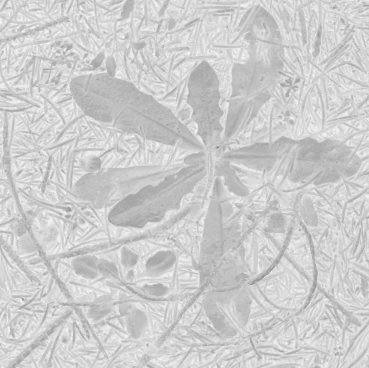} &
			\includegraphics[width=\imgW]{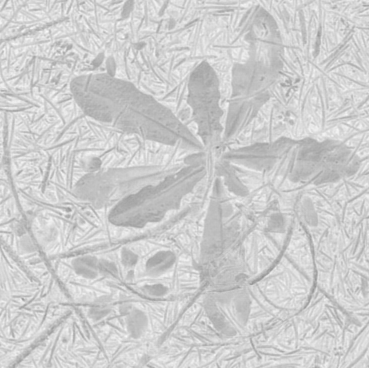} \\
			& \raisebox{\raiseboxheight}{\rotatebox{90}{mip 3}} &
			\includegraphics[width=\imgW]{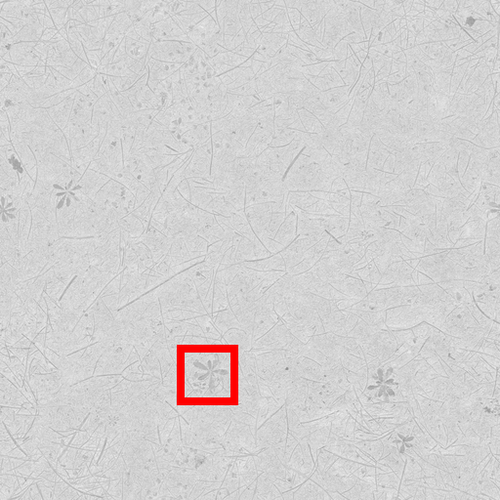} &
			\includegraphics[width=\imgW]{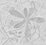} &
			\includegraphics[width=\imgW]{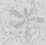 } &
			\includegraphics[width=\imgW]{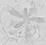} &
			\includegraphics[width=\imgW]{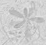} &
			\includegraphics[width=\imgW]{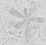}
		\end{tabular}
	}
	\caption{Comparison of different methods at 0.5~BPPC, where we selected to show
		a texture set for which NTC's PSNR was close to its average PSNR over all texture sets in our 20 texture evaluation dataset.
            Recall that neither AVIF nor JPEG XL provide random access to the texture data.
            For visualization purposes, the diffuse images were exposure compensated with factor $-1.0$ and
            tone mapped with ACES~\cite{Narkowicz2016}.
            Textures retrived from \url{https://kaimoisch.com/free-textures/}.
		}
	\label{fig_big_comparison_050}
\end{figure*}

\begin{figure*}[h!]
	\setlength{\tabcolsep}{1.0pt}%
	\renewcommand{\arraystretch}{0.7}
	\newcommand{\imgW}{0.13\textwidth}
	\newcommand{\linelen}{28.75mm}
        \newcommand{\raiseboxheight}{8.5mm}
	\centering
	{\footnotesize
		\begin{tabular}{ccccccc}
			& & & & \multicolumn{3}{c}{\line(1,0){\linelen}{\normalsize\  1.0 BPPC\ }\line(1,0){\linelen}} \\
			& & Original & Original & AVIF & JPEG XL & NTC \\
			\multirow{2}{*}[5mm]{\rotatebox{90}{{\normalsize diffuse map}}}
			& \raisebox{\raiseboxheight}{\rotatebox{90}{mip 0}} &
			\includegraphics[width=\imgW]{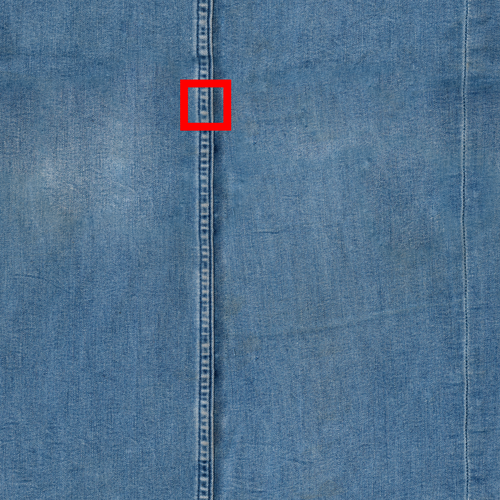} &
			\includegraphics[width=\imgW]{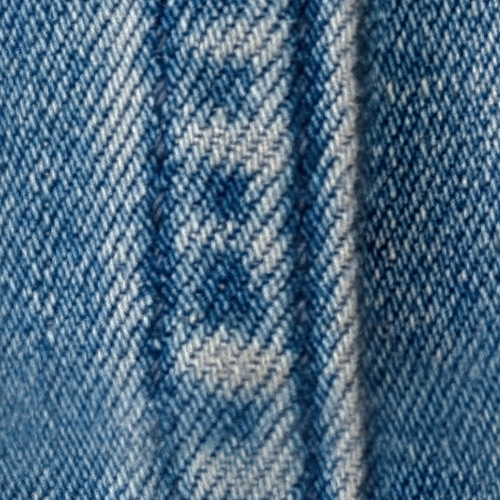} &
			\includegraphics[width=\imgW]{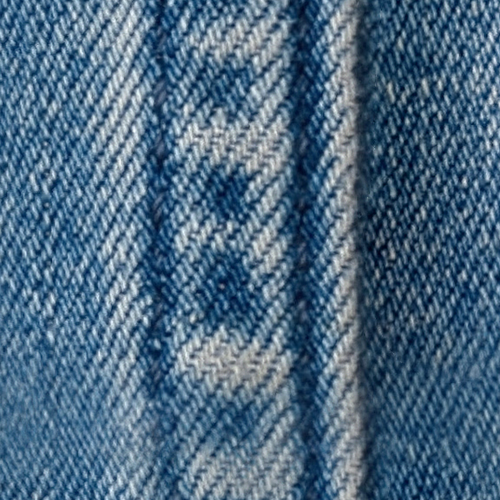} &
			\includegraphics[width=\imgW]{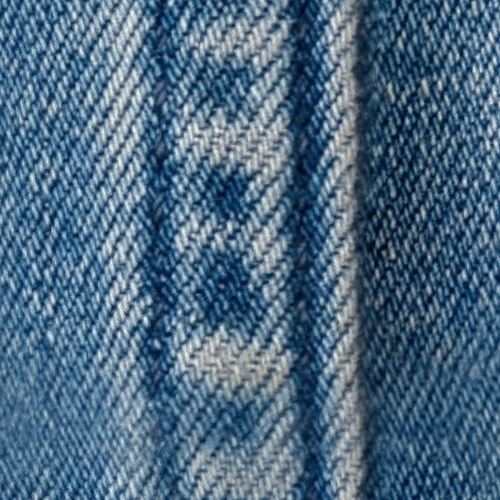} &
			\includegraphics[width=\imgW]{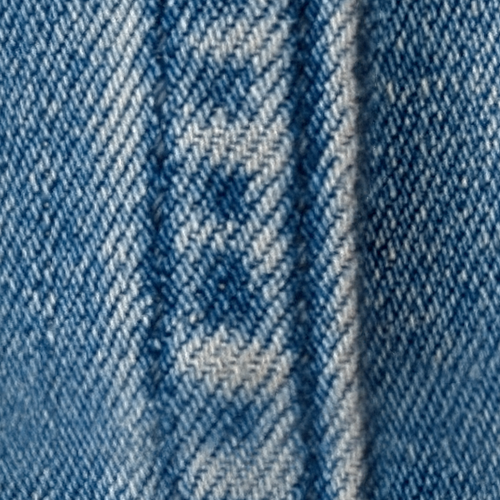} \\
			& \raisebox{\raiseboxheight}{\rotatebox{90}{mip 3}} &
			\includegraphics[width=\imgW]{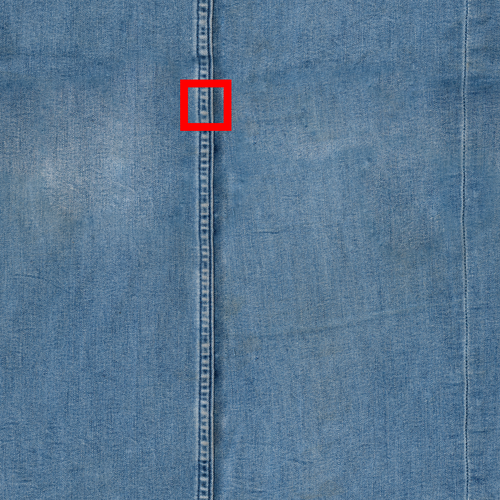} &
			\includegraphics[width=\imgW]{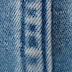} &
			\includegraphics[width=\imgW]{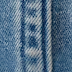} &
			\includegraphics[width=\imgW]{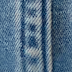} &
			\includegraphics[width=\imgW]{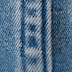} 
			\\
			\hline \vspace*{-1mm}
			\\
			\multirow{2}{*}[6mm]{\rotatebox{90}{{\normalsize normal map}}}
			& \raisebox{\raiseboxheight}{\rotatebox{90}{mip 0}} &
			\includegraphics[width=\imgW]{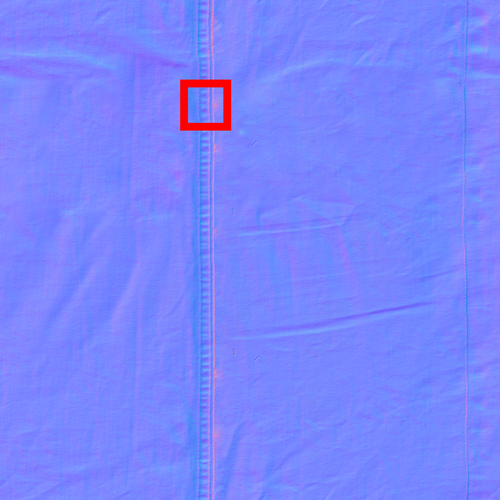} &
			\includegraphics[width=\imgW]{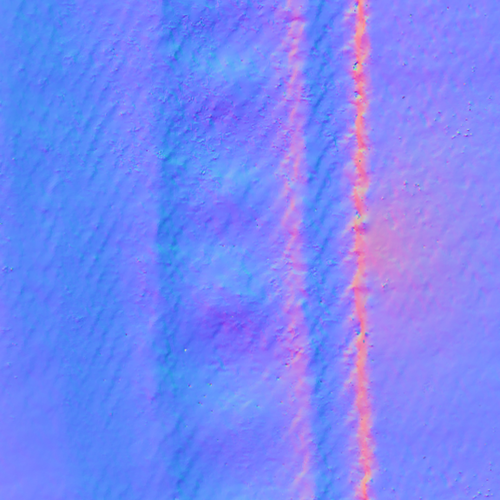} &
			\includegraphics[width=\imgW]{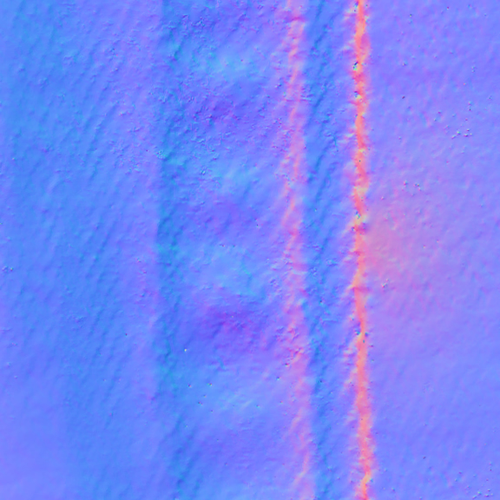} &
			\includegraphics[width=\imgW]{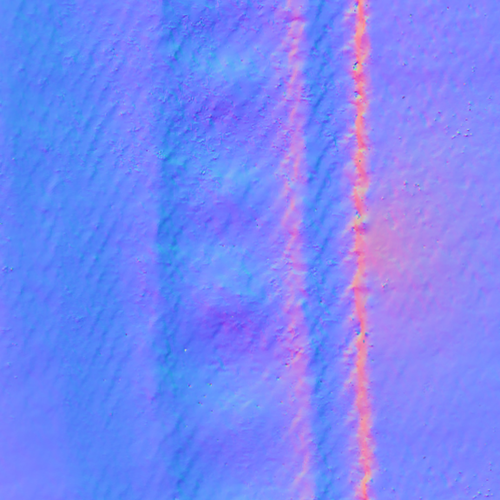} &
			\includegraphics[width=\imgW]{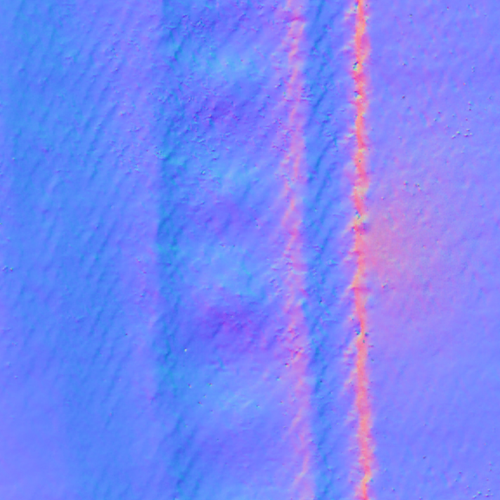} \\
			& \raisebox{\raiseboxheight}{\rotatebox{90}{mip 3}} &
			\includegraphics[width=\imgW]{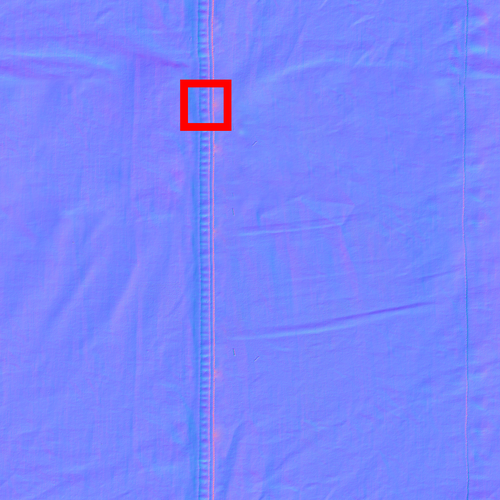} &
			\includegraphics[width=\imgW]{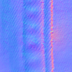} &
			\includegraphics[width=\imgW]{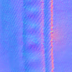} &
			\includegraphics[width=\imgW]{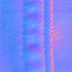} &
			\includegraphics[width=\imgW]{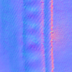}
			\\
			\hline \vspace*{-1mm}
			\\
			\multirow{2}{*}[10mm]{\rotatebox{90}{{\normalsize displacement map}}}
			& \raisebox{\raiseboxheight}{\rotatebox{90}{mip 0}} &
			\includegraphics[width=\imgW]{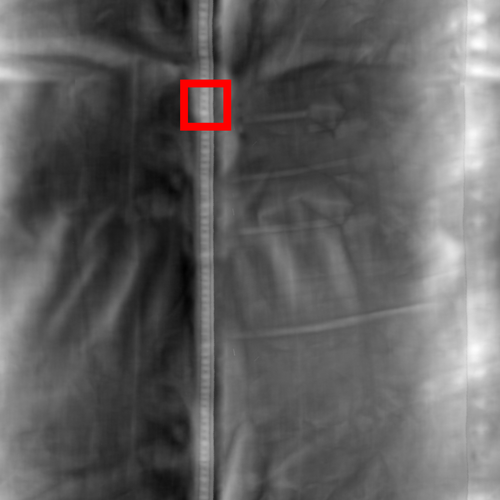} &
			\includegraphics[width=\imgW]{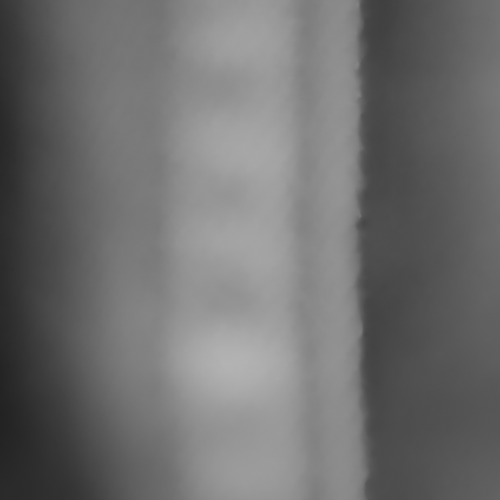} &
			\includegraphics[width=\imgW]{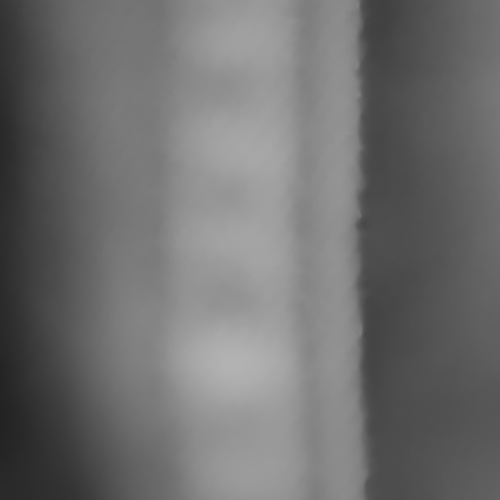} &
			\includegraphics[width=\imgW]{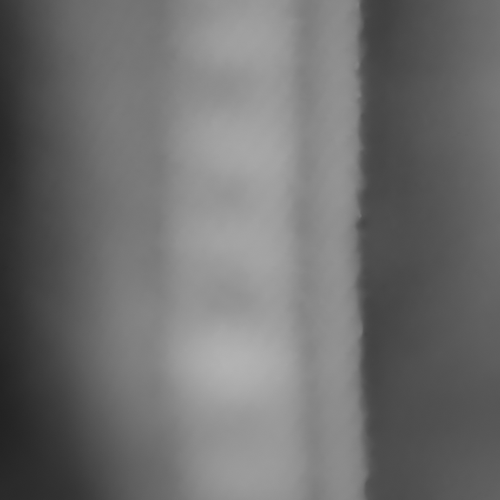} &
			\includegraphics[width=\imgW]{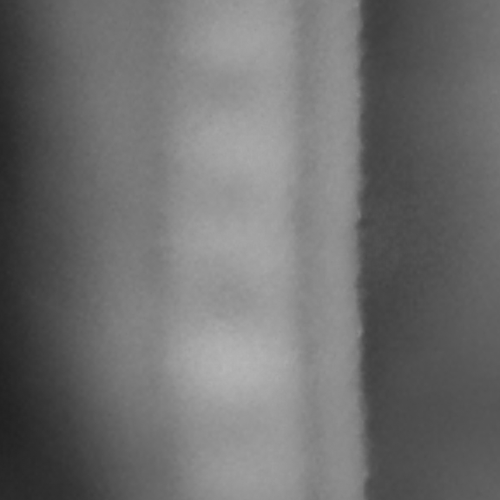} \\
			& \raisebox{\raiseboxheight}{\rotatebox{90}{mip 3}} &
			\includegraphics[width=\imgW]{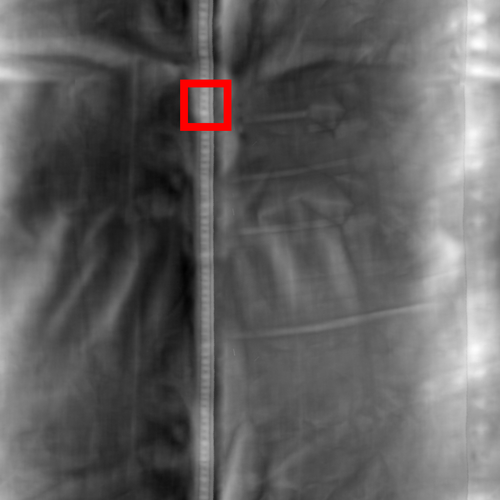} &
			\includegraphics[width=\imgW]{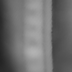} &
			\includegraphics[width=\imgW]{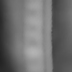} &
			\includegraphics[width=\imgW]{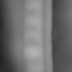} &
			\includegraphics[width=\imgW]{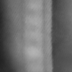}
			\\
			\hline \vspace*{-1mm}
			\\
			\multirow{2}{*}[2mm]{\rotatebox{90}{{\normalsize ARM}}}
			& \raisebox{\raiseboxheight}{\rotatebox{90}{mip 0}} &
			\includegraphics[width=\imgW]{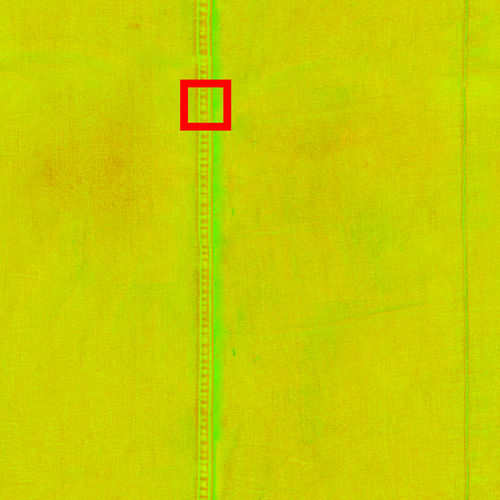} &
			\includegraphics[width=\imgW]{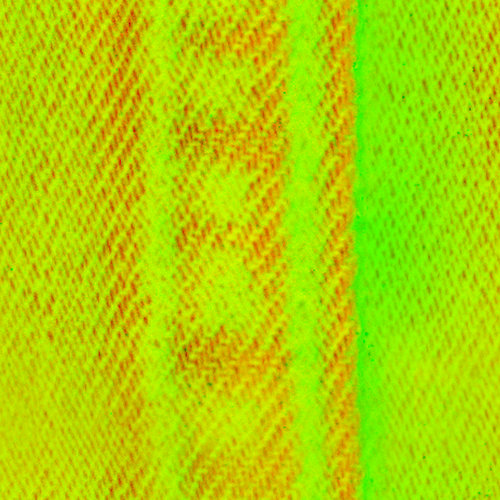} &
			\includegraphics[width=\imgW]{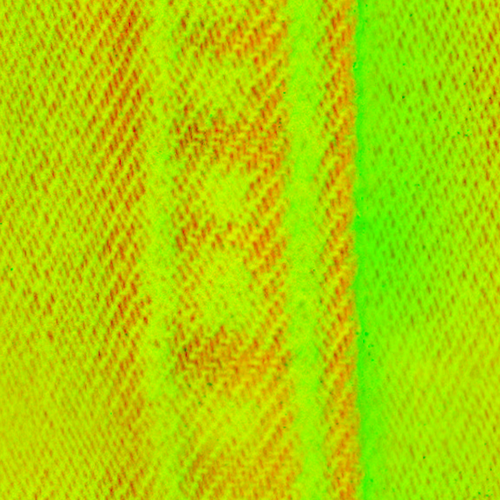} &
			\includegraphics[width=\imgW]{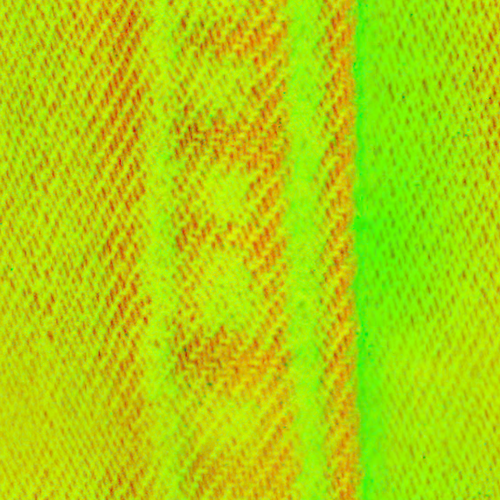} &
			\includegraphics[width=\imgW]{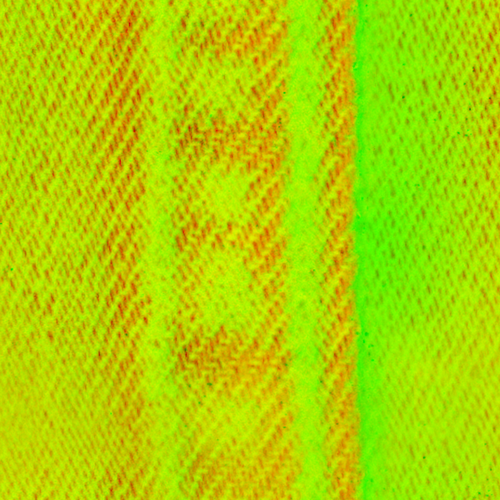} \\
			& \raisebox{\raiseboxheight}{\rotatebox{90}{mip 3}} &
			\includegraphics[width=\imgW]{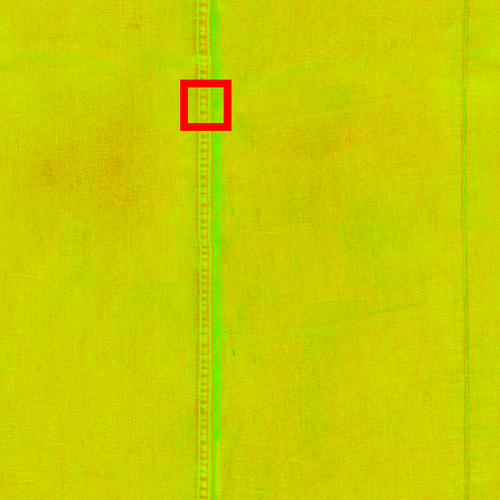} &
			\includegraphics[width=\imgW]{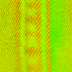} &
			\includegraphics[width=\imgW]{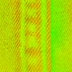} &
			\includegraphics[width=\imgW]{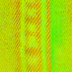} &
			\includegraphics[width=\imgW]{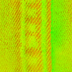}
		\end{tabular}
	}
	\caption{Comparison of different methods at 1.0~BPPC, where we selected to show
		a texture set for which NTC's PSNR was close to its average PSNR over all texture sets in our 20 texture evaluation dataset.
            Recall that neither AVIF nor JPEG XL provide random access to the texture data.
            Textures retrived from \url{https://polyhaven.com/}.
		}
	\label{fig_big_comparison_100}
\end{figure*}

Figures~\ref{fig_big_comparison_050} and \ref{fig_big_comparison_100} show results of texture compression
at other BPPC targets than the ones shown in the main paper. We note that the texture used to generate
Figure~\ref{fig_big_comparison_050} (\emph{Pine Forest Ground}) is half the size ($4096\times 4096$) of the texture
used to generate Figure~\ref{fig_big_comparison_100} (\emph{denim}) and the corresponding figure in the main paper.
In Figure~\ref{fig_big_comparison_050}, we compare to the medium-low-rate JPEG XL and AVIF configurations,
as well as the ASTC compressor (using tiles of size $12\times 12$, as those lead to a mean BPPC closer
to 0.50 for the evaluation data set compared to tiles of size $10\times 10$).
Figure~\ref{fig_big_comparison_100} compares the medium-bitrate compressors.
In the medium-bitrate case, it is difficult to spot any differences between the compressed textures
and the reference. For the medium-low-rate case, differences become visible, most notably for the
higher level mipmap where ASTC $12\times 12$ and AVIF show block artifacts. Here, NTC~0.5
show slight color changes in the diffuse texture. An added challenge of this texture set is that the normal map
is not aligned with the remaining textures.

\section{Comparison to Vector Quantization}
\label{sec:vq_comparison}

Vector quantization (VQ) is an alternate approach~\cite{oord2017} to discretizing the features, where each cell in a feature grid maps to an entry in a learned codebook or dictionary. 
During inference, the feature vectors can be replaced by a codebook index stored per grid cell. 
Unfortunately, the codebook size grows exponentially with the bitrate, making it prohibitively expensive to learn a codebook for higher quality levels. 
Therefore, in order to compare VQ with scalar quantization (SQ), we limit the size of the dictionary to 256 entries and assume multiple dictionaries, such that the overall storage size is the same. We only apply vector quantization to the higher resolution grid $G_0$, which is quantized to a smaller number of bits, while $G_1$ always uses scalar quantization with 12 channels and 4 bits.

Table~\ref{tab:sq_vs_vq} shows a comparison of SQ and VQ for a 4k texture using our lowest bitrate configuration (NTC~0.2) where each grid cell in $G_0$ stores 16 bits.
In the case of SQ, we use 8 channels which are quantized to 2 bits while in the case of VQ, we use two 256-entries dictionaries, which are referenced by two 8-bit indices respectively.
We compare SQ against two variants of VQ: VQ-8 has 8 channels per dictionary entry, which is similar to the size of the feature vector used in SQ, while VQ-16 uses 16 channels per dictionary entry and a correspondingly larger input layer in the decoder network.
The PSNR for all three quantization options are within 0.5~dB of each other. VQ-8 has slightly lower PSNR than SQ, while VQ-16 has a slightly higher PSNR, but with a higher cost for the input layer of the network. The training time for both VQ-8 and VQ-16 is more than 2.5$\times$ that of SQ. Given its simplicity and the significantly shorter training time, we choose scalar quantization for compression.

\section{Storage Cost}
\label{sec:storage_cost}
For completeness, we list the storage cost of NTC in Table~\ref{table:storage} for
our different compression profiles and for different texture set resolutions.

\begin{table*}[t]
\caption{Storage cost of NTC textures based on the compression profile and texture resolution. The network parameter size depends on the compression profile, but is constant for different texture sizes. Storage cost is independent of the input channel count.}
\vspace*{\tabvspace}
\resizebox{\textwidth}{!}{%
{\scriptsize
\begin{tabular}{@{}r|ccc|ccc|ccc|ccc@{}}
\toprule
 & \multicolumn{3}{c|}{NTC 0.2} & \multicolumn{3}{c|}{NTC 0.5} & \multicolumn{3}{c|}{NTC 1.0} & \multicolumn{3}{c}{NTC 2.25} \\ \midrule
Resolution & 2k$\times$2k & 4k$\times$4k & 8k$\times$8k & 2k$\times$2k & 4k$\times$4k & 8k$\times$8k & 2k$\times$2k & 4k$\times$4k & 8k$\times$8k & 2k$\times$2k & 4k$\times$4k & 8k$\times$8k \\ \midrule
NW (kB) & 24 & 24 & 24 & 27 & 27 & 27 & 25 & 25 & 25 & 27 & 27 & 27 \\
Grids (MB) & 0.875 & 3.5 & 14.935 & 2.125 & 8.5 & 36.269 & 4.25 & 17.0 & 72.534 & 9.5 & 38.0 & 162.135 \\
Total (MB) & 0.899 & 3.524 & 14.959 & 2.152 & 8.527 & 36.296 & 4.275 & 17.025 & 72.559 & 9.527 & 38.027 & 162.162 \\ \bottomrule
\end{tabular}
}
}
\label{table:storage}
\end{table*}

\section{Failed Experiments}
In the course of developing our method, we evaluated a few alternative methods for neural compression.
 We found that these methods, which are characterized by either increased complexity or 
 inferior quality, were unsuitable for the task of texture compression of material textures.
 We present these findings below.

\emph{Warped Grids.}
Prior work~\cite{lombardi2019neural} proposes to warp volumes with a non-uniform transformation for better resource utilization compared to a uniform grid.
We hypothesized that a similar approach, applied to images, could achieve some of the benefits of nonuniform bit allocation of entropy coding.
We found that the inclusion of warping grids led to an increase in PSNR scores between 0.1 and 0.9 dB, depending on the scale of the warping grid used.
However, after compressing and quantizing the warp grid, all the observed benefits could be achieved by simply allocating similar additional amount of storage to our latent grids. %

\emph{Nonuniform Quantization.}
We empirically observed that, prior to quantization, the distribution of our grid values  closely resembles a truncated normal distribution.
We tried adopting a normally distributed quantization scheme, but did not observe any quality improvement. We attribute this outcome to the fine-tuning of network weights after the freezing of the latent grids, which might compensate for the sub-optimal quantization distribution (see Section~\ref{quantization}).

\section{Usage of Other Compressors}
\label{supplement_compressors}
In this section, we describe which compressors we compare to and
their parameters.
Note that BCx and ASTC are specifically targeting texture compression/decompression
on GPUs and are designed to be random-access without entropy encoding.
JPEG XL and AVIF are more traditional image compressors and include
entropy encoding, which is a set of techniques that do not mesh well
with the random access requirement for textures. We have included them
still, since they are industry standards and because it may be worthwhile
to investigate how our method fares against such advanced techniques.
In fairness, it should be noted that neither JPEG XL nor AVIF were likely
designed to reach bitrates as low as 0.2~BPPC, which is NTC's lowest
target.
We have also used Basis/KTX2~\cite{Hurlburt2022}, which is part of the Khronos standard.
This format also uses entropy encoding, but during decompression, it can transcode
to many existing block-based texture compressions schemes, e.g., BCx, ETC, ASTC.

\subsection{BCx Compression}
For BCx compression~\cite{BCCompression2020}, we performed a smaller investigation
of existing tools, including 
AMD's Compressonator,\footnote{\url{https://gpuopen.com/compressonator/}} 
NVIDIA's Texture Tools,\footnote{\url{https://developer.nvidia.com/nvidia-texture-tools-exporter}}
and Intel's Fast ISPC Texture Compressor.\footnote{\url{https://github.com/GameTechDev/ISPCTextureCompressor}}
We used eight diffuse textures, eight normal maps, seven displacement maps, and seven roughness textures
for this evaluation. The diffuse, normal, and displacement maps were from PolyHaven, and the
roughness textures from ambientCG. The average PSNR for these three compressors 
over all textures were within $\pm 0.2$~dB.
While the ISPC texture compressor had the highest average score, we could not find a
command line tool version of it, and compressing a large number of high-resolution texture sets,
including mipmaps, manually
with their GUI was was prohibitively expensive.

For one-channel textures, e.g., roughness and displacement maps, AMD's tool wrote incorrect
output files, so we always used NVIDIA's tool for those. For the diffuse textures and normal
maps, AMD's tool produced slightly better result, so we used AMD's tool for those.
The highest parameter setting was used for NVIDIA's tool, while we used two refine steps
for BC1 (AMD) and quality 0.25 for BC7 (AMD). Going above those settings, mostly increased
compression times but not quality.

\subsection{ASTC Compression}
For ASTC~\cite{Nystad2012}, we used the texture tool
from ARM, who developed ASTC,\footnote{\url{https://github.com/ARM-software/astc-encoder}}
with the \texttt{-exhaustive} flag, which provided best quality.
Note that we used the two most aggressive variants of ASTC, which
compressed $12\times 12$ and $10\times 10$ tiles. All variants 
store 16 bytes per tile, so using $12\times 12$ tiles gives
$128\cdot 8 / (12\cdot 12)\approx 0.89$ bits per pixel for a three-channel texture.
Furthermore, note that, in Figure~\ref{fig:fig_quantitative2} in the main paper,
the ASTC results show average BPPC of around 0.5. This is a consequence
of storing BPPC over an entire texture set, which may include both
one- and three-channel textures. The same holds for other methods.

\begin{table}[t]
\newcolumntype{C}[1]{>{\centering\let\newline\\\arraybackslash\hspace{0pt}}m{#1}}
\centering
\caption{PSNR values with scalar quantization (SQ-8) and vector quantization (VQ-8, VQ-16) after optimization for 30k steps. }
\vspace*{\tabvspace}
\resizebox{\columnwidth}{!}{%
\begin{tabular}{@{}C{0.34\columnwidth}|C{0.34\columnwidth}|C{0.34\columnwidth}@{}}
\toprule
SQ-8 & VQ-8 & VQ-16 \\ \midrule
27.4 dB & 27.28 dB & 27.63 dB \\ \bottomrule
\end{tabular}%
}
\label{tab:sq_vs_vq}
\vspace*{-4mm}
\end{table}

\subsection{JPEG XL}
For JPEG XL~\cite{alakuijala2019jpeg}, we used the reference
implementation\footnote{\url{https://github.com/libjxl/libjxl}}
and its precompiled executables (v0.8.0) from November 2022.\footnote{\url{https://artifacts.lucaversari.it/libjxl/libjxl/latest/}}
We started by performing lossless compression, and if that
succeeded in reaching the target bitrate, our compression script exited.
Otherwise, our script performed a binary search to find the
quality setting that provided the sought-after bits per pixel per channel (BPPC).
To reduce compression times, we did an early-out if the compression rate
was within 2.5\% of the target compression rate.
We used the second highest value (8) for the effort parameter, since going to 9 (highest)
provided little to no additional quality, but further increased compression times.

\subsection{AVIF}
For AVIF~\cite{chen2018overview}, we used precompiled
executables using v0.11.1.\footnote{https://github.com/AOMediaCodec/libavif}
Similar to JPEG XL, our compression script for AVIF started
by attempting to do lossless compression and exited if that
reached the target compression rate. For all compression
with AVIF, we used the ``constant quality'' \texttt{-a end-usage=q} flag, since
this is common practice, and we also used quantization settings
\texttt{--min 0 --max 63}.
Next, our script performed a binary search on the
\texttt{-a cq-level} quantization parameter
\textit{without} chroma subsampling.
For very low bitrates, such 0.2~BPPC, this setting did not
always reach the target. In those cases, our script
continued with a new binary search with chroma subsampling
enabled (\texttt{--yuv 420}). In the end, the file
with the resulting bitrate closest to the target bitrate
was selected.
We used \texttt{--speed 3} since that resulted in reasonable compression
times and going lower did not substantially improve image quality.

\subsection{Basis/KTX2}
For Basis,\footnote{\url{https://github.com/BinomialLLC/basis_universal}} we
downloaded the code in early January 2023, and compiled it to use OpenCL for 
faster compression.
The flags we use for compression are:
\texttt{-opencl -ktx2 -uastc -uastc\_rdo\_l 1.0 -comp\_level 6},
where 6 offers the best image quality and takes the longest to compress.
All our results uses the file size of the output from the compressor.
For image quality, however, we unpacked the basis file and had
their decompressor transcode them to BC4 and BC7, respectively,
since those are the formats that we compare to in the main paper.

\begin{figure}[t]
\scalebox{0.5}{\input{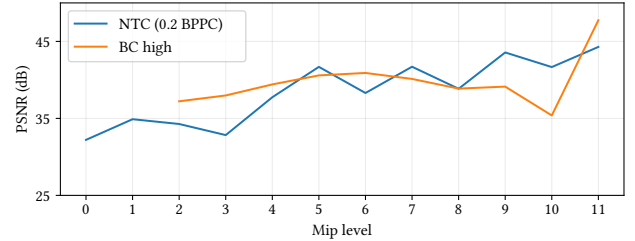}}
\caption{Comparison of iso-storage compression quality at each mip level between our algorithm and the high-rate BC configuration,
where we have used two fewer mipmap levels (0 and 1) to reach the same storage cost.}
\centering
\label{fig:mip_quality_plot}
\end{figure}

\section{Error and Quality Metrics on Texture Sets and Mip Chains}
\label{sec:error_computation}
For reproducibility and future research, this section documents how our quality and error metrics were computed
for texture sets and mip chains.

Given is a compressed texture set (including mip chains)
$\mathbf{T} = \left\{\mathbf{T}^0_0, \mathbf{T}^1_0, \mathbf{T}^2_0, \dots, \mathbf{T}^{M-1}_0, \dots, \mathbf{T}^0_{N-1}, \mathbf{T}^1_{N-1}, \mathbf{T}^2_{N-1}, \dots, \mathbf{T}^{M-1}_{N-1} \right\}$ with $N$ textures and $M$ mip levels, where mip $j$ of texture $i$ has resolution $w_j \times h_j \times c_i$. The values
in the textures are assumed to be in $[0,1]$.
For each mip level, $j$, we concatenate the corresponding textures, creating a tensor $\mathbf{T}^j$ of shape $w_j\times h_j\times c$, where $c = \sum_i c_i$.
The same concatenation is done for the reference texture set, $\mathbf{R}$, which contains the same number of textures as $\mathbf{T}$.

We compute the peak signal-to-noise ratio (PSNR) of $\mathbf{T}$ by summing the squared error over the mip chain and dividing by the total number of values in the tensor,
yielding its mean squared error (MSE). That is,
the PSNR is computed as
\begin{equation}
    \label{eq:texture_set_psnr}
    \textrm{PSNR}(\mathbf{R}, \mathbf{T}) = -10\log \left(\frac{\sum_j (\mathbf{R}^j - \mathbf{T}^j)^2}{c\sum_j w_j h_j}\right).
\end{equation}

Computing LPIPS~\cite{Zhang2018} for a texture set is a slightly more involved operation, as it requires 3-channel input. We consider a texture $\mathbf{T}_i^j$.
If it only has a single channel, we repeat its channels to give a 3-channel texture, $\mathbf{\tilde{T}}_i^j$.
If $\mathbf{T}_i^j$ has three channels,
we let $\mathbf{\tilde{T}}_i^j = \mathbf{T}_i^j$. We then normalize the texture so that its values are in $[-1,1]$, as is
required by LPIPS, creating $\mathbf{\bar{T}} = 2\mathbf{T}_i^j - 1$. The same procedure is used to create $\mathbf{\bar{R}}_i^j$.
Next, we compute LPIPS between the two tensors $\mathbf{\bar{T}}_i^j$ and $\mathbf{\bar{R}}_i^j$. LPIPS yields
an aggregate number $l_i^j=\textrm{LPIPS}\left(\mathbf{\hat{R}}_i^j, \mathbf{\hat{T}}_i^j\right)$
for the texture. We multiply the result by the number of channels $c_i$, effectively
making a 3-channel texture worth three times as much as a single-channel texture when computing LPIPS over the texture set.
Note that this is similar to the PSNR computations above. In addition, we multiply the result by the number of texels in the
texture. We again sum the results over all levels of the mip chain and divide by the total number of values in the texture set.
Formally, we have
\begin{equation}
    \label{eq:texture_set_lpips}
    \textrm{LPIPS}(\mathbf{R}, \mathbf{T}) = \frac{\sum_i \sum_j w_j h_j c_i l_i^j}{c \sum_j w_j h_j}.
\end{equation}
We note that LPIPS was computed using the \texttt{net='alex'} LPIPS model, as proposed by the authors of the metric.
Furthermore, LPIPS requires images that have resolutions of at least $32\times 32$. For textures smaller than
this, we apply zero-padding to make their sizes $32\times 32$.

The structural similarity index (SSIM)~\cite{Wang2004a} is a single-channel measure, providing a quality index between 0 and 1, where higher is better.
To instead report errors, we compute $1 - \textrm{SSIM}$. We get an SSIM value for
each channel in each texture in the mip chain. Similarly to the other two metrics, we weigh the result
by the number of texels on the current mip level and normalize by the number of values in the texture set to get the final result.
We get
\begin{equation}
    \label{eq:texture_set_ssim}
    1 - \textrm{SSIM}(\mathbf{R}, \mathbf{T}) =
    1 - \frac{\sum_i \sum_j w_j h_j c_i \textrm{SSIM}\left(\mathbf{R}_i^j, \mathbf{T}_i^j\right)}{c \sum_j w_j h_j},
\end{equation}
where $\textrm{SSIM}\left(\mathbf{R}_i^j, \mathbf{T}_i^j\right)$ computes SSIM for each channel in the texture and returns the average.
The SSIM computations include filtering with an $11\times 11$ Gaussian kernel. We do not apply the kernel for images that
are smaller than the kernel.

When we report LPIPS and SSIM errors over the entire data set, we take the mean of the errors computed for each texture set in the data set.
For PSNR, we compute the average of the per-texture MSE values retrieved during the computation of the PSNR values. The aggregate PSNR
is then computed using that average. Note that neither of these three aggregate error and quality values consider the
resolution of the textures in the texture set nor the total number of channels present within it. The effect of this
is that each texture in our diverse texture set has the same impact on the aggregate score.
Furthermore, note that the computations in Equations~\ref{eq:texture_set_psnr}-\ref{eq:texture_set_ssim} gives more weight to
the lower levels (i.e., higher resolutions) of the mip pyramid compared to the higher. The reasoning behind this is that the
lower levels will cover more pixels when used in rendered images.

Finally, we note that when we, for Figure~\ref{fig:mip_quality_plot}, compute the average quality for mip level $m$ over the entire data set,
we aggregate over all available mips at that level, independently of their resolution. Consider the aggregate error value for the second mip level, for example.
Despite the second mip level of a texture with resolution $1024\times 1024$ having resolution $512\times 512$ while the second mip level of a $4096\times 4096$ texture has resolution $2048\times 2048$, we add both of their error values into the aggregate for the second mip level, without weighting based on resolution.

\section{Evaluation Texture Set Details}
\label{sec_texture_set_selection}
The sizes of the textures range from $2048\times 2048$ to $8192\times 8192$.
The textures were either created by the authors or retrieved from
in-house or external sources. The external sources were:
ambientCG (\url{https://ambientcg.com/}), EISKO© (\url{https://www.eisko.com/}),
KaiMoisch (\url{https://kaimoisch.com/free-textures/}), and PolyHaven (\url{https://polyhaven.com/}).
Consisting of only a diffuse texture, the ``gradient 4k'' texture set
contains the fewest number of channels (three). This texture set
is the only one created by the authors and is provided as a difficult case. The ``Louise 4k'' set contains
the most channels (12 in total, consisting of diffuse: 3, normal: 3, roughness: 1, subsurface: 1, ambient occlusion: 1, displacement: 1, gloss: 1, and specular: 1).
Grayscale textures that were stored as RGB were converted to one-channel textures and
constant textures were removed.
EISKO and KaiMoisch provided 16-bit textures, which we converted to 8-bit before we used them.

The normal maps were originally provided as three-channel textures. We note that the
BC5 format is a two-channel format targeting normal maps, under the assumption
that the third component can be computed if the normals are of unit length.
Intially, we planned to convert all normal maps to two channels, but
found that the majority of the normal maps did not have normals of unit length.
Since the length of the normal sometimes is used to store another property
(for filtering normals, for example), we left them as three-channel
textures to avoid changing the intent of the texture artists.
However, since the quality of the normals is often extremely important,
we used the higher-quality BC7 format (instead of BC1).

We create the mipmap chains with a high-quality Lanczos downsampling
filter and stop when we have reached the level with a resolution of $4\times4$ pixels, since that is
the tile size of most block-based GPU compression schemes.

\end{document}